\documentclass[final,5p,times]{elsarticle}

\usepackage{amssymb, bm, amsmath,hyperref,multirow}
\usepackage{algorithm}

\usepackage{soul}

\usepackage[noend]{algpseudocode}
\usepackage{esint}

\algdef{SE}[SUBALG]{Indent}{EndIndent}{}{\algorithmicend\ }%
\algtext*{Indent}
\algtext*{EndIndent}

\usepackage{xcolor}

\newcounter{bla}

\journal{Computer Physics Communications}

\begin{document}

\begin{frontmatter}



\title{PyWolf: A PyOpenCL implementation for simulating the propagation of  partially coherent light}


\author[a,b,c]{Tiago E. C. Magalhães\corref{author}}

\author[a,b]{José M. Rebordão}

\cortext[author] {Corresponding author.\\\textit{E-mail address:} tecmagalhaes@fc.up.pt}
\address[a]{Departamento de Física, Faculdade de Ciências, Universidade de Lisboa, Edifício C8, Campo Grande, PT1749-016 Lisboa, Portugal}
\address[b]{Instituto de Astrof{\'i}sica e Ci{\^e}ncias do Espa\c{c}o, Edif{\'i}cio C8, Campo Grande, PT1749-016 Lisboa, Portugal}
\address[c]{Present address: Instituto de Física de Materiais Avançados, Nanotecnologia e Fotónica (IFIMUP), Departamento de Física e Astronomia, Faculdade de Ciências, Universidade do Porto, Rua do Campo Alegre s/n, 4169-007 Porto, Portugal}

\begin{abstract}
We present PyWolf, an open-source software capable of performing numerical simulations of partially coherent light propagation from two-dimensional light sources. PyWolf computes the evolution of a user-defined cross-spectral density function in the Fresnel and far field approximations, which enables the retrieval of second-order optical quantities of interest such as the spectral degree of coherence and spectral density for a given frequency. The open-source tool kit PyOpenCL is used to increase the computation speed. We present examples of propagation of different source models and optical systems to validate our implementation.  Performance results for the computation speed when using parallel computation through PyOpenCL is shown. Source models and propagation systems can be easily added to PyWolf, which has a graphical user interface built with PyQt5. This software can be of great utility for partially coherent light simulation problems that are difficult to treat analytically.

\end{abstract}


\end{frontmatter}



{\bf PROGRAM SUMMARY}

\begin{small}
\noindent
{\em Program Title:} PyWolf                                           \\
{\em Licensing provisions:} GPLv3                               \\
{\em Programming language:} Python                           \\
{\em Repository:} \url{https://github.com/tiagoecmagalhaes/PyWolf}           \\
{\em External routines:} NumPy, SciPy, Matplotlib, PyOpenCL, PyQt5         \\
{\em Nature of problem:} Numerical simulations of the propagation of partially coherent light from planar sources in the Fresnel or far field approximations. Computation time can be improved by using optimized versions of the Fast Fourier Transform (FFT) algorithm, but other calculations can only be increased through parallel computation.\\
{\em Solution method:} We use the open-source toolkit PyOpenCL to perform parallel computation on time-consuming calculations. The user can easily modify and add more features to the software, such as source models and propagation methods. A graphical user interface is also built using PyQt5 which enables the user to set input parameters, including the user's custom models, view the simulation results, export data, and save and load sessions. \\
   \\
\end{small}

\section{Introduction}
\label{sec:Introduction}

In second-order coherence theory, partially coherent light is described in the space-time domain by the mutual coherence function~\cite{Wolf1954_2} or in the space-frequency domain by the cross-spectral density function~\cite{Wolf1982}. In free space, these two quantities satisfy a pair of wave equations~\cite{Wolf1955,Parrent1959,Wolf1986a} (known as the Wolf equations). For propagation analysis purposes, it is usual to use the space-frequency approach, since the response of matter to light fields is, in most linear regime cases,  frequency-dependent~\cite{Wolf1982}. For planar sources, the integral solution of the Wolf equations yields a four-dimensional (4D) integral. In the Fresnel approximation~\cite{BornWolf1998,Yoshimori1995,Castaneda1997}, this integral reduces to two two-dimensional (2D) Fourier transforms, aside from the multiplication by the so-called propagators or kernels.

If the cross-spectral density function is known, we can extract optical quantities such as the spectral density and the correlation between pairs of points (i.e., the spectral degree of coherence). In general, the computational array describing the cross-spectral density function at a given angular frequency is 4D~\cite{Magalhaes2017}, and complex values must be stored for each pair of 2D points. As the size of this 4D array increases, the amount of system memory increases as well as the computation time. For instance, consider two 4D arrays with sizes $N^4$ and $(N+1)^4$. The difference between the number of elements in both arrays is $4N^3+6N^2+4N+1$. If $N=100$, the difference between both arrays is approximately $4\times10^6$ elements. Although highly optimized algorithms for solving 2D Fourier transforms exist, such as the Fast Fourier Transform (FFT) algorithm, we are still left with other computational tasks that require a large computation time, such as building the initial source model into an array and multiplying by the propagators. 

Several techniques and approximations have been developed in the past, not only to decrease the computation time but also to simplify the analytical calculations. For example, the coherence-mode decomposition method introduced by Emil Wolf~\cite{Wolf1981} reduces the calculation to two-dimensional (2D) integrals. It consists of writing the cross-spectral density as a superposition of contributions from completely
coherent modes~(for a detailed description, see reference~\cite{ostrovsky2006coherent}). For instance, perfectly coherent light is characterized by one mode and as the spectral degree of coherence "decreases", the number of modes increases. To find these modes, we need to find solutions to a 2D equation rather than a 4D one. Burvall et al.~\cite{Burvall2009} developed a method, known as the elementary function method, which is easier to handle numerically when compared to the coherent-mode decomposition. However, the source cross-spec\-tral density cannot be arbitrary and must satisfy several assumptions (e.g., cross-spectral density must be real). Several other methods that involve approximations on the source model were proposed (see reference~\cite{Burvall2009} and references therein). In this work, we are interested in a general approach since we want to use arbitrary source models, where the propagation is limited only in terms of the Fresnel approximation and single-frequency description. To deal with the huge computation time required to perform operations on 4D arrays, we implemented an algorithm that uses parallel computing devices to decrease the computation time, the results of which have already been published~\cite{Magalhaes2017,Magalhaes2019thesis}. Although our initial problems are related to light propagation and imaging, its formal mathematical structure can be found elsewhere, for example in~\cite{MandelWolf1995,Goodman2015} 

In this work, we present an open-source software written in the Python language capable of propagating the cross-spectral density function using parallel computing devices in very time-consuming tasks. To achieve this goal, we used PyOpenCL~\cite{Kloeckner2012}, a tool kit that gives access through Python to the Open Computing Language (OpenCL), a programming language developed by the Khronos Group to support heterogeneous computing \cite{Kaeli2015}. To enable a user-friendly environment and easy customization of source models and propagation functions, we added a graphical user interface using the PyQt5 package. PyWolf aims at providing researchers a framework for faster simulations of partially coherent light propagation from user-defined source models. Researchers in different fields (e.g., coherence theory, optical instrumentation, and astrophysics) may find this tool important when dealing with the propagation of partially coherent light in optical systems of interest.

This article is structured as follows. In Section~\ref{sec:Theory} we will briefly review the theory of the propagation of partially coherent light from planar sources in the space-frequency domain using the cross-spectral density function. In Section~\ref{sec:numeric}, we will present the data and numerical model for the propagation of the cross-spectral density function and discuss the PyOpenCL implementation. In Section~\ref{sec:pywolf} we discuss PyWolf's framework and graphical user interface. In Section~\ref{sec:examples}, we present simulation examples to validate our implementation. In Section~\ref{sec:performance}, we discuss performances in terms of computation time with and without the use of PyOpenCL in the computational tasks. Finally, in Section~\ref{sec:conclusion}, we conclude and discuss future improvements.

\section{Theory}
\label{sec:Theory}

In this section, we will summarize the theory underlying  the propagation of the cross-spectral density function from the source plane to the observation plane in free space. For a more detailed analysis of coherence theory and propagation of partially coherent light, we refer the reader to references \cite{MandelWolf1995,Wolf2007,Goodman2015}.

\subsection{Propagation in a homogeneous and isotropic medium}
\label{subsec:freesace}
\begin{figure}[!t]
\centering
\includegraphics[width=0.49\textwidth]{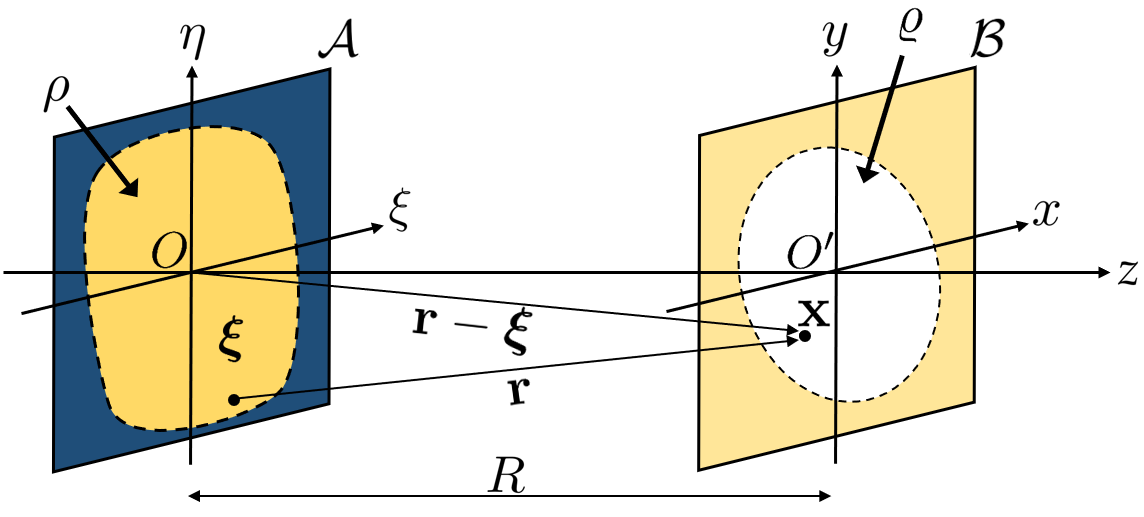}
\caption{\label{fig:prop} (Color online) Illustration of the notation used for the propagation of the source cross-spectral density function. $O$ and $O'$ are the origins of planes $\mathcal{A}$ and $\mathcal{B}$, respectively.}
\end{figure}

With reference to Fig.~\ref{fig:prop}, let $W_\mathcal{A}(\boldsymbol{\xi}_1,\boldsymbol{\xi}_2,\omega)$ be the cross-spectral density function of plane $\mathcal{A}$, containing  a source $\rho$, given by
\begin{equation}
    W_{\mathcal{A}}(\boldsymbol{\xi}_1,\boldsymbol{\xi}_2,\omega) = \sqrt{S_{\mathcal{A}}(\boldsymbol{\xi}_1,\omega)} \sqrt{S_{\mathcal{A}}(\boldsymbol{\xi}_2,\omega)} \,\mu_{\mathcal{A}}(\boldsymbol{\xi}_1,\boldsymbol{\xi}_2,\omega)\,,
\label{eq:W_def}
\end{equation}
where $\boldsymbol{\xi}_1=(\xi_1,\eta_1)$ and $\boldsymbol{\xi}_2=(\xi_2,\eta_2)$ are a pair of source points, $S_{\mathcal{A}}(\boldsymbol{\xi}_i,\omega)$ is the spectral density at point $\boldsymbol{\xi}_i$, and $\mu_{\mathcal{A}}(\boldsymbol{\xi}_1,\boldsymbol{\xi}_2,\omega)$ is the spectral degree of coherence. Both quantities can be extracted from the cross-spectral density function $W_{\mathcal{A}}(\boldsymbol{\xi}_1,\boldsymbol{\xi}_2,\omega)$ using their definitions~\cite{Wolf2007}:
\begin{equation}
    S_{\mathcal{A}}(\boldsymbol{\xi},\omega)= W_{\mathcal{A}}(\boldsymbol{\xi},\boldsymbol{\xi},\omega)\,,\label{eq:defS}
\end{equation}
\begin{equation}
    \mu_{\mathcal{A}}(\boldsymbol{\xi}_1,\boldsymbol{\xi}_2,\omega)= \frac{W_{\mathcal{A}}(\boldsymbol{\xi},\boldsymbol{\xi},\omega)}{\sqrt{S_{\mathcal{A}}(\boldsymbol{\xi}_1\,\omega)}\sqrt{S_{\mathcal{A}}(\boldsymbol{\xi}_2,\omega)}}\,.\label{eq:defMu}
\end{equation}
While the spectral density of the source $\rho$ may be rather arbitrary, the spectral degree of coherence is limited to the non-negative definiteness constraint of the cross-spectral density function~\cite{Gori2007,Mei2013}:
\begin{equation}
    \iiiint f^{*}(\boldsymbol{\xi}_1)f(\boldsymbol{\xi}_2)W(\boldsymbol{\xi}_1,\boldsymbol{\xi}_2,\omega)\,\mathrm{d}\xi_1\,\mathrm{d}\eta_1\,\mathrm{d}\xi_2\,\mathrm{d}\eta_2\geq0\,.
\end{equation}

Let $\mathcal{B}$ be an observation plane parallel to plane $\mathcal{A}$ and distanced by $R$.
In the framework of second-order coherence theory, the propagation of the cross-spectral density function $W(\boldsymbol{\xi}_1,\boldsymbol{\xi}_2,\omega)$ to plane $\mathcal{B}$ is given by~\cite{WolfCarter1978,Yoshimori1995}
\begin{align}
W_{\mathcal{B}}(\mathbf{x}_{1},\mathbf{x}_{2},\omega) = & \iiiint   h^{*}(\mathbf{x}_1,\boldsymbol{\xi}_{1},\omega)\,W_{\mathcal{A}}(\boldsymbol{\xi}_{1},\boldsymbol{\xi}_{2},\omega) \nonumber \\
  &\times \,h(\mathbf{x}_2,\boldsymbol{\xi}_{2},\omega)\,\mathrm{d}\xi_{1}\,\mathrm{d}\eta_{1}\,\mathrm{d}\xi_{2}\,\mathrm{d}\eta_{2}\,,
\label{eq:W_general}
\end{align}
where the integration is taken over plane $\mathcal{A}$, the asterisk $*$ denotes the complex conjugate and $h(\mathbf{x},\boldsymbol{\xi},\omega)$ is the response function, i.e., the propagation kernel from point $\mathbf{x}$ to point $\boldsymbol{\xi}$. In free space, $h(\mathbf{x},\boldsymbol{\xi},\omega)$ is given by~\cite{Wolf1955,WolfCarter1978}
\begin{equation}
    h(\mathbf{x},\boldsymbol{\xi},\omega) = \frac{\Lambda}{r} \exp\left( \mathrm{i}kr\right)\,,\label{eq:h_free}
\end{equation}
where $\mathrm{i}$ is the imaginary unit, $k$ is the wavenumber, $r=|\mathbf{r}|$ is the distance between points $\boldsymbol{\xi}$ and $\mathbf{x}$ (see Fig. \ref{fig:prop}), and $\Lambda$ is the obliquity factor \cite{Goodman2005}, which is different depending on the type of source in plane $\mathcal{A}$. If source $\rho$ is primary (e.g., the Sun), the obliquity factor assumes the simple form of $\Lambda=1/4\pi$. If the source is secondary (e.g., an aperture), it can be shown that $\Lambda$ is approximately given by \cite{Yoshimori1995}
\begin{equation}
    \Lambda \approx \frac{k}{4\pi \mathrm{i}} \left(\cos\theta_i+\cos\theta_o\right)\,,
\end{equation}
where 
\begin{equation}
    \cos\theta_i = \frac{k_z^{(i)}}{|\mathbf{k}^{(i)}|}\,,
\end{equation}
\begin{equation}
    \cos\theta_o = \frac{\left(\mathbf{r}-\boldsymbol{\xi} \right)_z}{\left|\mathbf{r}-\boldsymbol{\xi}  \right|}\,,
\end{equation}
$\mathbf{k}^{(i)}=k^{(i)}\,\mathbf{r}^{(i)}/r^{(i)}$ is the incident wave vector on the secondary source and $k_z^{(i)}$ denotes the $z$-component of the vector $\mathbf{k}^{(i)}=(k_x^{(i)},k_y^{(i)},k_z^{(i)})$, in the same direction of the optical axis $z$ (see reference~\cite{Yoshimori1995} for more details). Let $\varrho$ be a finite region in plane $\mathcal{B}$ such that the distance $R$ is larger than the characteristic scale of this region. For instance, if $\varrho$ is a circular region with radius $a_\varrho$, we assume that $R\gg a_\varrho$. If we consider that points $\mathbf{x}_1$ and $\mathbf{x}_2$ are within region $\varrho$, we may approximate the term $\Lambda/r$ of equation~(\ref{eq:h_free}) as
\begin{equation}
    \frac{\Lambda}{r} \approx \frac{\Lambda}{R}\,.\label{eq:approx_L}
\end{equation}
Henceforth, we will always use this approximation. Substituting equation~(\ref{eq:h_free}) into~(\ref{eq:W_general}) and using the Fresnel approximation, the cross-spectral density at plane $\mathcal{B}$ can be written as \cite{Yoshimori1995}
\begin{align}
W_{\mathcal{B}}(\mathbf{x}_{1},\mathbf{x}_{2},\omega)\simeq & \,C\,q^{*}(R,\mathbf{x}_{1},\omega)\,q(R,\mathbf{x}_{2},\omega)\iint q^{*}(R,\boldsymbol{\xi}_{1},\omega) \, \nonumber \\
    &\times \exp\left(-2\pi\mathrm{i}\mathbf{K}_{1}\cdot\boldsymbol{\xi}_{1}\right) \mathrm{d}\xi_{1}\,\mathrm{d}\eta_{1} \iint\,W_{\mathcal{A}}(\boldsymbol{\xi}_{1},\boldsymbol{\xi}_{2},\omega)\,\nonumber \\
    &\times q(R,\boldsymbol{\xi}_{2},\omega)\exp\left(-2\pi\mathrm{i}\mathbf{K}_{2}\cdot\boldsymbol{\xi}_{2}\right)\,\mathrm{d}\xi_{2}\,\mathrm{d}\eta_{2}\,,
\label{eq:W_fresnel}
\end{align}
where $C$ is a constant defined as 
\begin{equation}
    C = \frac{\Lambda_1^{*} \Lambda_2}{R^2}\,,\label{eq:C}
\end{equation}
$\mathbf{K}$ is the spatial frequency of the 2D Fourier transform given by
\begin{equation}
    \mathbf{K} = \frac{\omega}{2 \pi c R}\mathbf{x}\,,\label{eq:K}
\end{equation}
and $q(R,\boldsymbol{\xi},\omega)$ is a function defined as
\begin{equation}
q(R,\boldsymbol{\xi},\omega)=\exp\left(\mathrm{i}\frac{\omega}{2cR}\boldsymbol{\xi}^{2}\right)=\exp\left(\mathrm{i}\pi\frac{\boldsymbol{\xi}^{2}}{\lambda R}\right)\,,\label{eq:q_factor}
\end{equation}
where $\boldsymbol{\xi}^2=\boldsymbol{\xi}\cdot\boldsymbol{\xi}$ and we used the fact that, in free space, $\omega=2\pi c/\lambda$. The following properties can be easily derived from equation~(\ref{eq:q_factor}):
\begin{equation}
q\left(R,\frac{\boldsymbol{\xi}}{\beta},\omega\right)= q\left(\beta^2 R,\boldsymbol{\xi},\omega\right)\,,\label{eq:qprop1}
\end{equation}
\begin{equation}
q(R,\boldsymbol{\xi}_2,\omega)\,q^{*}(R,\boldsymbol{\xi}_1,\omega)= q\left( R,\sqrt{\boldsymbol{\xi}^2_2-\boldsymbol{\xi}^2_1},\omega\right) \,,\label{eq:qprop1}
\end{equation}
\begin{equation}
q(R_1,\boldsymbol{\xi},\omega)\,q^{*}(R_2,\boldsymbol{\xi},\omega)= q\left( \frac{R_1 R_2}{R_2-R_1},\boldsymbol{\xi},\omega\right) \,,\label{eq:qprop2}
\end{equation}
\begin{equation}
q(R,\boldsymbol{\xi},\omega)\,q^{*}(R,\boldsymbol{\xi},\omega)= 1 \,,\label{eq:qprop3}
\end{equation}
where $R_1$ and $R_2$ are arbitrary distances and $\beta$ is a constant (e.g., a transverse magnification).

If the distance $R$ is large enough compared with the source $\rho$ dimensions, we can use the far field approximation~\cite{Gori2005}, i.e., set $q(R,\boldsymbol{\xi}_{1,2},\omega)\approx 1$, and equation~(\ref{eq:W_fresnel}) reduces to
\begin{align}
W_{\mathcal{B}}(\mathbf{x}_{1},\mathbf{x}_{2},\omega) = &\,C\,q\left(R,\sqrt{\mathbf{x}_{2}^2-\mathbf{x}_{1}^2},\omega\right) \iint \exp\left(-\mathrm{i}\frac{\omega}{cR}\mathbf{x}_{1}\cdot\boldsymbol{\xi}_{1})\right)  \nonumber \\
& \times \mathrm{d}\xi_{1}\,\mathrm{d}\eta_{1} \iint W_{\mathcal{A}}(\boldsymbol{\xi}_{1},\boldsymbol{\xi}_{2},\omega) \nonumber \\
&\times \exp\left(-\mathrm{i}\frac{\omega}{cR}\mathbf{x}_{2}\cdot\boldsymbol{\xi}_{2}\right)\,\mathrm{d}\xi_{2}\,\mathrm{d}\eta_{2}\,,
\label{eq:W_farfield}
\end{align}
where we used equation~(\ref{eq:qprop1}) to simplify the product of the two $q$ functions. For the sake of simplicity, we will hereafter refer to these functions as propagators. Notice that, apart from the propagators, the far field cross-spectral density is the 2D Fourier transform of the cross-spectral density in plane $\mathcal{A}$, which is expected from the van Cittert-Zernike theorem (see, for example, Chapter 5 of reference~\cite{,MandelWolf1995}). 

After calculating $W_\mathcal{B}(\mathbf{x}_1,\mathbf{x}_2,\omega)$ in equation~(\ref{eq:W_fresnel}) [or equation~(\ref{eq:W_farfield}), depending on the approximation], we can use it to extract second-order optical quantities of interest in plane $\mathcal{B}$, such as the spectral density $S_{\mathcal{B}}(\mathbf{x},\omega)$ and the spectral degree of coherence $\mu_{\mathcal{B}}(\mathbf{x}_1,\mathbf{x}_2,\omega)$, using equations~(\ref{eq:defS}) and~(\ref{eq:defMu}), respectively.

\subsection{Imaging System}
We have just reviewed the propagation of the cross-spectral density function from the source plane to an observation plane, in free space. We can use the same procedure in more complex systems. We will briefly address the case of an ideal imaging system, which will be used to validate our numerical model. A more detailed analysis can be found, for example, in reference~\cite{Castaneda1997}. 

\begin{figure}
\centering
\includegraphics[width=0.49\textwidth]{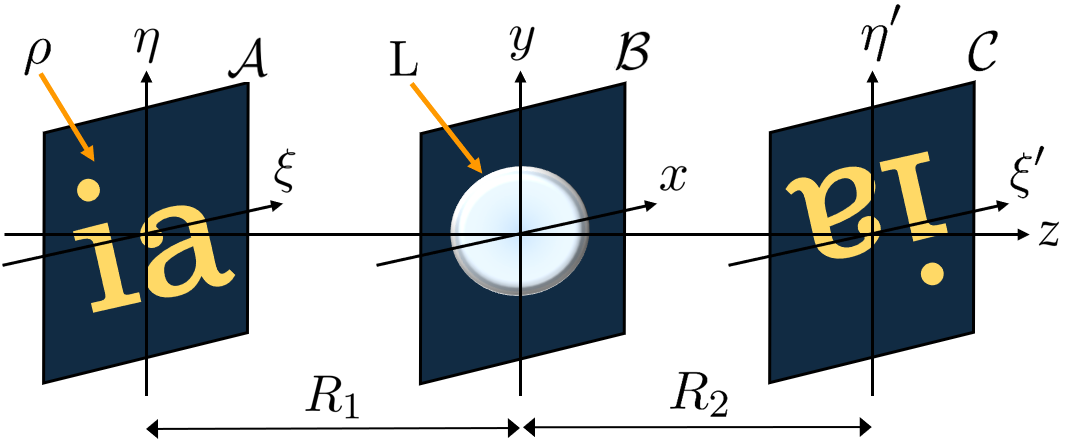}
\caption{\label{fig:imaging} (Color online) Illustration of the notation used for the propagation of the cross-spectral density function in an imaging system. $\mathrm{L}$ is a positive lens.}
\end{figure}

Let $\mathcal{A}$ be once again the source plane, as represented in Fig.~\ref{fig:imaging}, containing a primary source $\rho$. A positive thins lens, L, with focal length $f$ is placed in plane $\mathcal{B}$, distanced by $R_1$. Since we are dealing with an ideal imaging system with one thin lens, the entrance and exit pupil will lie in the same plane, $\mathcal{B}$. The cross-spectral density $W_{\mathcal{B}}^{(-)}(\mathbf{x}_{1},\mathbf{x}_{2},\omega)$ immediately before plane $\mathcal{B}$ is calculated using the Fresnel approximation of equation~(\ref{eq:W_fresnel}), and we can write it as
\begin{align}
W_{\mathcal{B}}^{(-)}(\mathbf{x}_{1},\mathbf{x}_{2},\omega)= & \,C_1^2\,\,\mathcal{F}_{\boldsymbol{\xi}_2}^{-1} \left\{ \mathcal{F}_{\boldsymbol{\xi}_1} \left\{ q\left(R_1,\sqrt{\boldsymbol{\xi}_{2}^2-\boldsymbol{\xi}_{1}^2},\omega\right) \right. \right. \nonumber \\
    \times &\left. \left. W_{\mathcal{A}}(\boldsymbol{\xi}_{1},\boldsymbol{\xi}_{2},\omega) \right\}_{\mathbf{x}_1} \right\}_{\mathbf{x}_2} q\left(R_1,\sqrt{\mathbf{x}_{2}^2-\mathbf{x}_{1}^2},\omega\right)\,,
\label{eq:WA_imaging}
\end{align}
where $C_1=(\Lambda_1^{*}\Lambda_2)/{R_1^2}$. The symbols $\mathcal{F}_{\boldsymbol{\xi}}\left\{\right\}_{\mathbf{x}}$ and $\mathcal{F}_{\boldsymbol{\xi}}^{-1}\left\{\right\}_{\mathbf{x}}$ denote the direct and inverse 2D Fourier transform, respectively, with respect to the spatial variable $\boldsymbol{\xi}$, and the subscript $\mathbf{x}$ represents the new spatial domain, i.e.,
\begin{equation}
    \hat{f}(\mathbf{x})= \mathcal{F}_{\boldsymbol{\xi}}\left\{f(\boldsymbol{\xi}) \right\}_{\mathbf{x}}=\iintop_{-\infty}^{+\infty} f(\boldsymbol{\xi}) \exp{\left(-\frac{\mathrm{i}\,\omega}{c} \mathbf{x}\cdot\boldsymbol{\xi} \right)}\,\mathrm{d}\xi \,\mathrm{d}\eta\,,
\end{equation}
where $f(\boldsymbol{\xi})$ is the function to be Fourier transformed and $\hat{f}(\mathbf{x})$ is its Fourier transform in the new domain $\mathbf{x}$.
Immediately after plane $\mathcal{B}$, the cross-spectral density $W_{\mathcal{B}}^{(+)}(\mathbf{x}_{1},\mathbf{x}_{2},\omega)$ is given by~\cite{Yoshimori1995}  
\begin{equation}
    W_{\mathcal{B}}^{(+)}(\mathbf{x}_{1},\mathbf{x}_{2},\omega)=T^{*}(\mathbf{x}_1,\omega)\,W_{\mathcal{B}}^{(-)}(\mathbf{x}_{1},\mathbf{x}_{2},\omega)\,T(\mathbf{x}_2,\omega)\,,
\end{equation}
where $T(\mathbf{x},\omega)$ is the transmission function of L, given by
\begin{equation}
    T(\mathbf{x},\omega)=P(\mathbf{x})\,q^{*}(f,\mathbf{x},\omega)\,,
\end{equation}
where $q^{*}(f,\mathbf{x},\omega)$ is defined in equation~(\ref{eq:q_factor}) and $P(\mathbf{x})$ is the pupil function. The latter can be written as~\cite{Castaneda1997} 
\begin{equation}
    P(\mathbf{x}) = |P(\mathbf{x})|\mathrm{e}^{\mathrm{i}\Phi (\mathbf{x})}\,,
\end{equation}
where $\Phi(\mathbf{x})$ is the aberration function (see, for example, reference~\cite{BornWolf1998}). 

Let $\mathcal{C}$ be an imaging plane, distanced by $R_2$ from plane $\mathcal{B}$. The cross-spectral density $W_{\mathcal{C}}(\boldsymbol{\xi}_{1}',\boldsymbol{\xi}_{2}',\omega)$ at plane $\mathcal{C}$ is calculated, once again, using the Fresnel approximation of equation~(\ref{eq:W_fresnel}). Using the lens law
\begin{equation}
    \frac{1}{R_1}+\frac{1}{R_2}-\frac{1}{f} = 0\,,\label{eq:imaging}
\end{equation}
and using the properties of the $q$ function of equations~(\ref{eq:qprop1})-(\ref{eq:qprop3}), we obtain 
\begin{align}
W_{\mathcal{C}}(\boldsymbol{\xi}_{1}',\boldsymbol{\xi}_{2}',\omega)= & \,C_1\,C_2\,q\left(R_2,\sqrt{\boldsymbol{\xi}_{2}'^{2}-\boldsymbol{\xi}_{1}'^{2}},\omega\right)    \nonumber \\
    \times & \mathcal{F}_{\mathbf{x}_2}^{-1} \left\{ \mathcal{F}_{\mathbf{x}_1} \left\{  \,\mathcal{F}_{\boldsymbol{\xi}_2}^{-1} \left\{ \mathcal{F}_{\boldsymbol{\xi}_1} \left\{ q\left(R_1,\sqrt{\boldsymbol{\xi}_{2}^2-\boldsymbol{\xi}_{1}^2},\omega\right) \right. \right. \right. \right.\nonumber \\
    \times &\left. \left. \left. \left. W_{\mathcal{A}}(\boldsymbol{\xi}_{1},\boldsymbol{\xi}_{2},\omega) \right\}_{\mathbf{x}_1} \right\}_{\mathbf{x}_2} P'(\mathbf{x}_1,\mathbf{x}_2)\right\}_{\boldsymbol{\xi}_{1}'} \right\}_{\boldsymbol{\xi}_{2}'} \,,
\label{eq:WC_imaging}
\end{align}
where $C_2=(\Lambda_1^{*}\Lambda_2)/R_2^2$ and $P'(\mathbf{x}_1,\mathbf{x}_2)$ is defined as
\begin{equation}
    P'(\mathbf{x}_1,\mathbf{x}_2) = P^{*}(\mathbf{x}_1)\,P(\mathbf{x}_2)\,.
\end{equation}
If we consider an aberration free imaging system ($\Phi=0$) and neglect the finite extend of the lens L, we can set $P(\mathbf{x}_{1,2})=1$ (see, for example, reference~\cite{Goodman2005}). Furthermore, if the lens L is placed at a distance $R_1$ such that 
\begin{equation}
    q\left(R_1,\sqrt{\boldsymbol{\xi}_{2}^2-\boldsymbol{\xi}_{1}^2},\omega\right)\approx 1\,,
\end{equation}
the cross-spectral density at plane $\mathcal{C}$ will be simply
\begin{align}
   W_{\mathcal{C}}(\boldsymbol{\xi}_{1}',\boldsymbol{\xi}_{2}',\omega) = C'\,q\left(R_2,\sqrt{\boldsymbol{\xi}_{2}'^{2}-\boldsymbol{\xi}_{1}'^{2}},\omega\right) W_{\mathcal{A}}(\boldsymbol{\xi}_{1}',\boldsymbol{\xi}_{2}',\omega)\,,
\end{align}
where $C' = C_1 C_2$. The spectral density $S_{\mathcal{C}}$ will thus be
\begin{equation}
    S_{\mathcal{C}}(\boldsymbol{\xi}',\omega) = C'\,S_{\mathcal{A}}(\boldsymbol{\xi}',\omega)\,,
\end{equation}
an expected result since planes $\mathcal{A}$ and $\mathcal{C}$ are conjugated. However, we must keep in mind that we have neglected the finite extent of the lens L and aberrations, otherwise, we would not have a perfect mapping from plane $\mathcal{A}$ to the image plane.

\section{Numerical Propagation of the Cross-Spectral Density using PyOpenCL}
\label{sec:numeric}

We will now discuss the data representation for the cross-spectral density function for a single frequency. We will then describe the numerical implementation of the propagation of the cross-spectral density in free space.

\subsection{Data representation}
\label{sec:num_model}

\begin{figure}
\centering
\includegraphics[width=0.46\textwidth]{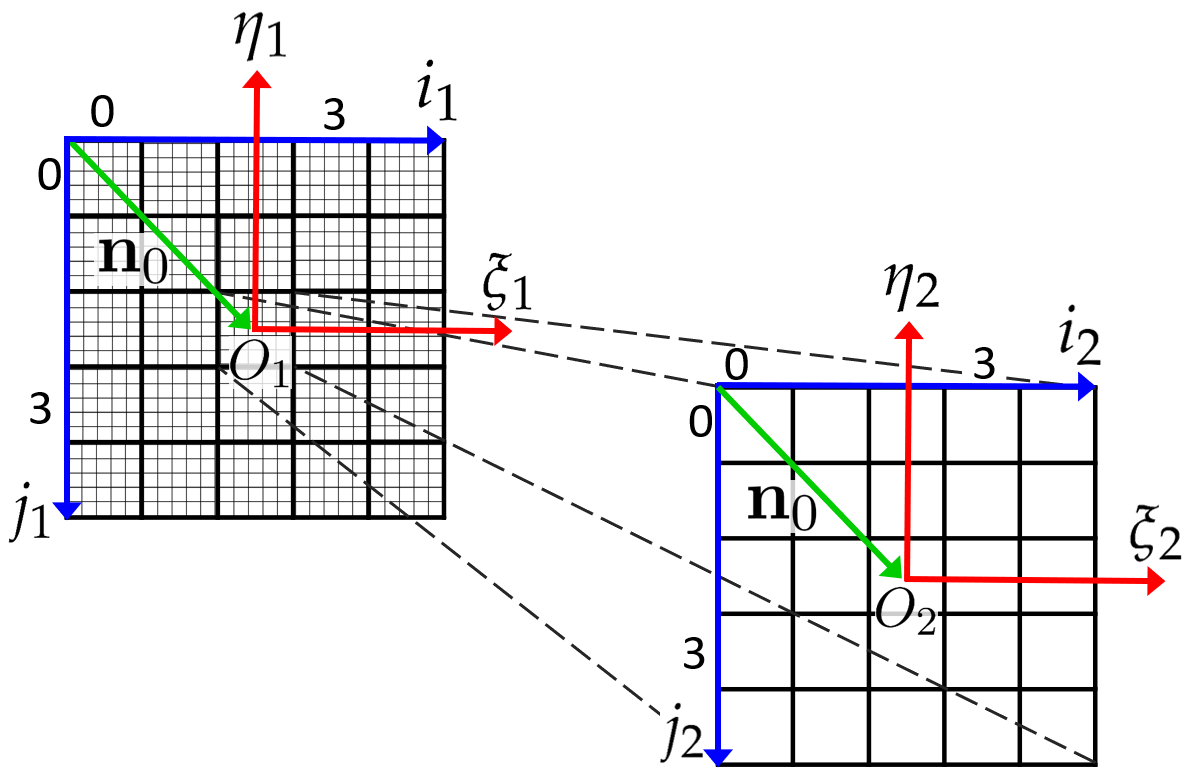}
\caption{\label{fig:CSDM} (Color online) Illustration of a  CSDA with $N=5$. $i_1$ and $j_1$ are the typical indexes of an array, in this case, $i_{1,2},j_{1,2}\in\{0,1,2,4\}$. $\xi_{1,2},\eta_{1,2}$ are the modified indexes.}
\end{figure}

Let $W_{\mathcal{A}}(\boldsymbol{\xi}_1,\boldsymbol{\xi}_2,\omega_0)$ be the cross-spectral density function, at a given angular frequency $\omega _0$, of plane $\mathcal{A}$, which contains a light source $\rho$, as illustrated in Fig. \ref{fig:prop}. Let also $\mathbf{W}^{(0)}$  be a $N\times N\times N\times N$ array, representing a sampling of $W_{\mathcal{A}}(\boldsymbol{\xi}_1,\boldsymbol{\xi}_2,\omega_0)$ in a finite squared-size portion of plane $\mathcal{A}$.  The superscript $(0)$ denotes the initial plane from which we want to propagate the cross-spectral density function. We will denote $\mathbf{W}^{(0)}$ as the cross-spectral density array (CSDA) instead of the cross-spectral density matrix to avoid confusion from the electric cross-spectral density matrix found in the unified theory of polarization and coherence (e.g., see Chapter 9 of reference~\cite{Wolf2007}). Computationally, the position vectors $\boldsymbol{\xi}_1$ and $\boldsymbol{\xi}_2$ are defined as
\begin{equation}
    \boldsymbol{\xi}_{1,2} = (\xi_{1,2},\eta_{1,2}) =\Delta\xi \left( \mathbf{n}_{1,2}-\mathbf{n}_0 \right)\,,\label{eq:point}
\end{equation}
where $\Delta \xi$ is the spatial resolution of the CSDA (spatial sampling) and $\mathbf{n}_{1,2}$ and $\mathbf{n}_0$ are two adimensional vectors, respectively defined as
\begin{equation}
    \mathbf{n}_{1,2} = (i_{1,2},-j_{1,2})\,\,\,\,(i_{1,2},j_{1,2}=0,1,...,N-1)\,,\label{eq:n12}
\end{equation}
\begin{equation}
    \mathbf{n}_0 = \left( \frac{N}{2},-\frac{N}{2} \right)\,,
\end{equation}
where $i$ and $j$ are indexes of the CSDA, which correspond to the indexes of a typical numeric array. Figure \ref{fig:CSDM} illustrates an example of a CSDA for $N=5$, which has a total size of 625, i.e., $N^4$.

Let us define, for the sake of brevity, $M=N/2$. Since the CSDA $\mathbf{W}^{(0)}$ is a 4D array, it can be described as
\begin{equation}
\mathbf{W}^{(0)}=\left(\begin{array}{ccc}
\mathbf{W}_{0,0}^{(0)} & \cdots & \mathbf{W}_{0,M}^{(0)}\\
\vdots & \ddots & \vdots\\
\mathbf{W}_{M,0}^{(0)} & \cdots & \mathbf{W}_{M,M}^{^{(0)}}
\end{array}\right)\,,
\end{equation}
where $\mathbf{W}_{i_{1},j_{1}}^{(0)}\,(i_1,j_1=0,1,...,N-1)$ is a $N\times N$ array for a given point $\boldsymbol{\xi}_1=\Delta\xi(\mathbf{n}_1-\mathbf{n}_0)$. Therefore, we can write $\mathbf{W}_{i_{1},j_{1}}^{(0)}$ as
\begin{equation}
\mathbf{W}_{i_{1},j_{1}}^{(0)}=\left(\begin{array}{ccc}
\mathbf{W}_{i_{1},j_{1};0,0}^{(0)} & \cdots & \mathbf{W}_{i_{1},j_{1};0,M}^{(0)}\\
\vdots & \ddots & \vdots\\
\mathbf{W}_{i_{1},j_{1};M,0}^{(0)} & \ldots & \mathbf{W}_{i_{1},j_{1};M,M}^{(0)}
\end{array}\right)\,,
\end{equation}
where $\mathbf{W}_{i_{1},j_{1};i_2,j_2}^{(0)}$ is a complex number which can be sampled from the cross-spectral density function, i.e.,
\begin{equation}
    \mathbf{W}_{i_{1},j_{1},i_2,j_2}^{(0)} = W\left[(\mathbf{n}_1-\mathbf{n}_0)\Delta\xi,(\mathbf{n}_2-\mathbf{n}_0)\Delta\xi,\omega_0\right]\,.
\end{equation}
Thus, for a given source model, which is described by the spectral density and the spectral degree of coherence [see equation~(\ref{eq:W_def})], we must construct a given CSDA $\mathbf{W}^{(0)}$, where each value $\mathbf{W}_{i_{1},j_{1},i_2,j_2}^{(0)}$ will be sampled from the cross-spectral density function of such model. 

\subsection{Creation of the CSDA $\mathbf{W}^{(0)}$ with PyOpenCL}
\label{sec:sourceCSD}

To start the simulation, PyWolf requires an initial CSDA $\mathbf{W}^{(0)}$. To define this and other arrays, PyWolf uses the NumPy Python library, namely, the \textit{ndarray} object. 
The initial CSDA $\mathbf{W}^{(0)}$ can be created according to the selected model in PyWolfit or uploaded to PyWolf through the \textit{npy} format. In the latter case, $\mathbf{W}^{(0)}$ can be custom-built by the user or obtained by a previous simulation of PyWolf. If not uploaded, PyWolf will create this CSDA according to what the user-defined in the input parameters of PyWolf (see Section~\ref{sec:pywolf}), which we will next describe.

The first step is to define an "empty" $N\times N\times N\times N$ array and initialize it with zeros. Then, depending on the user input, we can add a geometrical model (e.g., a circle) by filling elements with ones (see, for example, Algorithm~\ref{alg:circle} in~\ref{sec:GSMS_example}). For instance, if at a given point source $\boldsymbol{\xi}=\Delta\xi(i,j)$ light is not present, then we have that $\mathbf{W}^{(0)}_{i,j,i,j}=0+0\,\mathrm{i}$. Otherwise, if light is present, we have that $\mathbf{W}^{(0)}_{i,j,i,j}=1+0\,\mathrm{i}$.
If we do not wish to apply any geometry, the initial CSDA is initialized with ones, rather than zeros.

After modeling the geometry, we must apply a given source model to the CSDA. In general, this involves the computation of $N^4$ elements in $\mathbf{W}^{(0)}$, a heavy time-consuming task as we increase $N$. To decrease the computing time, PyOpenCL is used to compute each $N\times N$ matrix  $\mathbf{W}^{(0)}_{i_1,j_1}$. Four loops (\textit{for} cycles) are used to go through $N^4$ elements. To increase the speed, we can use PyOpenCL to parallelize the last two loops. We will use the same procedure whenever we have to deal with four imbricated loops. 

Mathematically, the source model in plane $\mathcal{A}$ can be described as
\begin{equation}
    W_{\mathcal{A}}(\boldsymbol{\xi}_1,\boldsymbol{\xi}_2,\omega_0) = Y_\mathcal{A}(\boldsymbol{\xi}_1)\,Y_\mathcal{A}(\boldsymbol{\xi}_2)\,W_{\rho}(\boldsymbol{\xi}_1,\boldsymbol{\xi}_2,\omega_0)\,,\label{eq:GSM_source}
\end{equation}
where $Y_\mathcal{A}(\boldsymbol{\xi})$ is a function defining the geometry in plane $\mathcal{A}$, consisting  of only ones and zeros, and $W_{\rho}(\boldsymbol{\xi}_1,\boldsymbol{\xi}_2,\omega_0)$ is the cross-spectral density function of a given model. An example of the creation of the CSDA  $\mathbf{W}^{(0)}$ for a Gaussian Schell-model source is described in~\ref{sec:GSMS_example}.

\subsection{Spatial sampling and resolution}

Computationally, the spatial resolution $\Delta x$ in the observation plane $\mathcal{B}$ will be determined by the 2D Fourier transforms of equation~(\ref{eq:W_fresnel}) or, in the far field approximation, of equation~(\ref{eq:W_farfield}). In either case, the spatial frequency vector $\mathbf{K}$ of each 2D Fourier transform is given by equation~(\ref{eq:K}). If we want to sample the observation plane with a spatial step $\Delta x$, the spatial frequency step $\Delta K$ is, according to equation~(\ref{eq:K}), given by
\begin{equation}
    \Delta K = \frac{\omega}{2\pi c R}\Delta x\,. \label{eq:DeltaK_theory}
\end{equation}
If we use the FFT algorithm to solve each Fourier transform, the spatial frequency resolution $\Delta K_{\mathrm{FFT}}$ is given by \cite{Rao2011}
\begin{equation}
    \Delta K_{\mathrm{FFT}} = \left( N \Delta \xi \right)^{-1}\,.\label{eq:DeltaKFFT}
\end{equation}
Thus, equalizing equations~(\ref{eq:DeltaK_theory}) and~(\ref{eq:DeltaKFFT}), we find that
\begin{equation}
    \Delta x = \frac{2\pi cR}{\omega N \Delta\xi} \,.\label{eq:spatial_resolution}
\end{equation} 
Note that if we use zero padding~\cite{Rao2011} to perform the FFT, we must consider the total size of the 2D matrix in the term $N$, even if, later, we decide to crop the matrix to its original size. The relation between the spatial resolutions  $\Delta \xi$ and $\Delta x$ in equation~(\ref{eq:spatial_resolution}) is used whenever we propagate the CSDA from one plane (with spatial resolution $\Delta \xi$) to another ($\Delta x$), separated by $R$.

\subsection{Propagation}
\label{sec:prop}

We will now discuss the numerical implementation of the propagation of an arbitrary CSDA $\mathbf{W}^{(0)}$, for a given single frequency $\omega_0$, according to the theory described in Section \ref{subsec:freesace} in terms of the propagation between the source plane and the observation plane. Propagation through multiple planes, such as in imaging, is straightforward and follows the same steps, as long as the Fresnel approximation is valid. PyOpenCL is used to parallelize computational tasks that are time-consuming.  Algorithm~\ref{alg:steps} describes all the steps taken during the propagation, starting in the creation of the source CSDA $\mathbf{W}^{(0)}$ and ending with the computation of the final CSDA $\mathbf{W}^{(6)}$ at the observation plane.

\begin{algorithm}[b!]
\caption{All steps used for propagation in free space of for a given model.}\label{alg:steps}
\textbf{Input:} $N\times N \times N \times N$ matrix $\mathbf{W}^{(0)}$, array size integer $N$, spatial sampling $\Delta \xi$, central frequency $\omega_0$, distance $R$, speed of light $c$, Boolean value $\mathrm{usePyOpenCL}$, geometry parameters $\mathrm{GeoPars}$, source model parameters, $\mathrm{SourcePars}$ \\
\textbf{Output:} $N\times N \times N \times N$ matrix $\mathbf{W}^{(6)}$ \\
\begin{algorithmic}[1]
\State $\mathbf{W}^{(0)} \gets \textsc{zeros}(N,N,N,N)$ \Comment{\textcolor{blue}{Creates a 4D complex array filled with zeros.}}
\State $\mathbf{W}^{(0)} \gets \textsc{geometry}(\mathbf{W}, N, \mathrm{GeoPars}, \text{usePyOpenCL})$ \phantom N \phantom N \phantom N  \Comment{\textcolor{blue}{Builds the geometry of the source plane (e.g., function \textsc{Circle} of Algorithm~\ref{alg:circle}}).}
\State $\mathbf{W}^{(0)} \gets \textsc{model}(\mathbf{W}, N, \mathrm{SourcePars}, \text{usePyOpenCL})$ \phantom N \phantom N \phantom N \phantom N \Comment{\textcolor{blue}{Builds the coherence model for the CSDA (e.g., function \textsc{GSMsource} of Algorithm~\ref{alg:GSM}).}}
\State $\mathbf{W}^{(6)} \gets \textsc{freespaceprop}$($\mathbf{W}^{(0)}$, $N$, $\Delta\xi$, $\omega_0$, $R$, $c$, $\text{farfield}$, $\text{usePyOpenCL}$)
\end{algorithmic}
\end{algorithm}
\begin{algorithm}[!t]
\caption{Propagation function in free space for a given CSDA $\mathbf{W}^{(0)}$.}\label{alg:prop}
\begin{algorithmic}[1]
\Function{freespaceprop}{${\mathbf{W}^{(0)}}$, $N$, $\Delta\xi$, $\omega_0$, $R$, $c$, $\text{farfield}$, $\text{usePyOpenCL}$}
    \If{$\text{farfield}=\text{False}$}
    \State $\mathbf{W}^{(1)}\gets \textsc{qfunction}$($\mathbf{W}^{(0)}, N, \Delta\xi, \omega_0, R, c$)
    \Else 
    \State $\mathbf{W}^{(1)} \gets \mathbf{W}^{(0)}$
    \EndIf

    \For{$i_1=0$ to $N-1$ \textbf{step} 1}
          \For{$j_1=0$ to $N-1$ \textbf{step} 1}	    \State $\mathbf{W}_{i_1,j_1}^{(2)}\gets \mathrm{FFT2}\left(\mathbf{W}_{i_1,j_1}^{(1)}\right)$  \Comment{\textcolor{blue}{2D direct FFT.}} 
   \EndFor
       \EndFor 
    \State $\mathbf{W}^{(3)} \gets \textsc{transpose}(\mathbf{W}^{(2)},2,3,0,1)$ \Comment{\textcolor{blue}{Swapping axes 2 and 3 with 0 and 1, respectively.}}
    
    \For{$i_2=0$ to $N-1$ \textbf{step} 1}
          \For{$j_2=0$ to $N-1$ \textbf{step} 1}	    \State $\mathbf{W}_{i_2,j_2}^{(4)}\gets \mathrm{iFFT2}(\mathbf{W}_{i_2,j_2}^{(3)})$ \Comment{\textcolor{blue}{2D inverse FFT.} }
   \EndFor
       \EndFor 

    \State $\mathbf{W}^{(5)} \gets \textsc{transpose}(\mathbf{W}^{(4)},2,3,0,1)$ \Comment{\textcolor{blue}{Swapping axes 2 and 3 with 0 and 1, respectively.}}

\State $\Delta x \gets 2\pi c R/(\omega_0 N \Delta\xi)$
 \State $\mathbf{W}^{(6)}\gets \textsc{qfunction}(\mathbf{W}^{(5)}, N, \Delta x, \omega_0, R, c)$   
     \State \Return  $\mathbf{W}'$
\EndFunction

\end{algorithmic}
\end{algorithm}
\begin{algorithm}[t]
\caption{Propagator $q$ function.}\label{alg:q_func}
\begin{algorithmic}[1]

\Function{qfunction}{${\mathbf{W}},N, {\Delta\xi}, {\omega_0}, {R}, c,\mathrm{usePyOpenCL}$}

    \For{$i_1=0$ to $N-1$ \textbf{step} 1}
      \For{$j_1=0$ to $N-1$ \textbf{step} 1}
      \State$\boldsymbol{\xi}_1=\left(i_1-\frac{N}{2},\frac{N}{2}-j_1 \right)\Delta\xi$
        
        \If{$\mathrm{usePyOpenCL}=\mathrm{True}$}
            \State \For{ $i_2,j_2=0$ to $N-1$ \textbf{step} 1} \textbf{in parallel}
                \Indent
                \State$\boldsymbol{\xi}_2\gets\left(i_2-M,M-j_2 \right)\Delta\xi$
     \State $\mathbf{W}_{i_1,j_1;i_2,j_2}' \leftarrow \mathbf{W}_{i_1,j_1;i_2,j_2}$ \phantom N \phantom N \phantom N \phantom N \phantom N  $\times\,\exp \left[ \mathrm{i}\frac{\omega_0}{2cR}\left(\boldsymbol{\xi}_2^2-\boldsymbol{\xi}_1^2 \right) \right]  $                               
                \EndIndent
                \EndFor
    \Else 
    \EndIf 
    \For{$i_2=0$ to $N-1$ \textbf{step} 1}
      \For{$j_2=0$ to $N-1$ \textbf{step} 1}
        \State$\boldsymbol{\xi}_2=\left(i_2-\frac{N}{2},\frac{N}{2}-j_2 \right)\Delta\xi$
     \State $\mathbf{W}_{i_1,j_1;i_2,j_2}' \leftarrow \mathbf{W}_{i_1,j_1;i_2,j_2}$ \phantom N \phantom N \phantom N \phantom N \phantom N  $\times\,\exp \left[ \mathrm{i}\frac{\omega_0}{2cR}\left(\boldsymbol{\xi}_2^2-\boldsymbol{\xi}_1^2 \right) \right]  $   

      \EndFor
    \EndFor
      \EndFor
    \EndFor
    
     \State \Return  $\mathbf{W}'$
\EndFunction	
\end{algorithmic}
\end{algorithm}

After creating the source CSDA $\mathbf{W}^{(0)}$ (lines 1-3 of Algorithm~\ref{alg:prop}), we use equation~(\ref{eq:W_fresnel}) to propagate it to the observation plane, which is described in Algorithm~\ref{alg:prop} by the function $\textsc{freespaceprop}$. If the far field approximation is not used,
the CSDA $\mathbf{W}^{(0)}$ will be multiplied by the propagators ($q$ functions) and we will obtain the CSDA $\mathbf{W}^{(1)}$. For this procedure, function $\textsc{qfunction}$ is used (see Algorithm~\ref{alg:q_func}) and the computation time can be decreased if we use PyOpenCL, which parallelizes two of the four loops used to compute all elements of the CSDA. If the far field approximation is used, this step is skipped and, therefore, $\mathbf{W}^{(1)}=\mathbf{W}^{(0)}$.

The next step is to compute $2N^2$ 2D Fourier transforms. For this task, we will use the Fast Fourier Transform (FFT) algorithm. First, $N^2$ 2D direct Fourier transforms in the $\boldsymbol{\xi}_1$ domain (lines 6-8 of Algorithm~\ref{alg:prop}) are computed. When completed, we end up with the CSDA $\mathbf{W}^{(2)}$. Then, we need to swap axes 2 and 3 (corresponding to the $\boldsymbol{\xi}_2$ domain of the CSDA $\mathbf{W}^{(2)}$) with axes 0 a 1 (corresponding to the $\boldsymbol{\xi}_1$ domain). This transposed CSDA is now denoted as $\mathbf{W}^{(3)}$. Then, we perform $N^2$ inverse 2D Fourier transforms in the $\boldsymbol{\xi}_2$ domain (lines 10-12 of Algorithm~\ref{alg:prop}), obtaining the CSDA $\mathbf{W}^{(4)}$. After that, we perform the same transpose as before in the CSDA $\mathbf{W}^(4)$, switching axes 2 and 3 with 0 and 1 (lines 13 of Algorithm~\ref{alg:prop}) and we obtain the new CSDA $\mathbf{W}^{(5)}$. Note that there is no need to perform this transpose since the cross-spectral density function is Hermitian~\cite{MandelWolf1995}, i.e.,
\begin{equation}
    W(\mathbf{x}_2,\mathbf{x}_1,\omega) = W^{*}(\mathbf{x}_1,\mathbf{x}_2,\omega)\,.
\end{equation}
It is not a time-consuming task and we leave it for the sake of clarity. After computation of all 2D Fourier transforms, we need to calculate the spatial resolution  $\Delta x$ of equation~(\ref{eq:spatial_resolution}). We are left to multiply the CDSA $\mathbf{W}^{(5)}$ by the $q$ function again (line 15 of Algorithm~\ref{alg:prop}). The final CSDA is denoted by $\mathbf{W}^{(6)}$.

\subsection{Retrieval of the spectral density and spectral degree of coherence}
\label{sec:retrieve}

After computing the CSDA $\mathbf{W}^{(6)}$, we can extract optical quantities for a given frequency $\omega_0$ such as the spectral density $S(\mathbf{x},\omega_0)$ and the spectral degree of coherence $\mu(\mathbf{x}_1,\mathbf{x}_2,\omega_0)$. In the former case, since we are only dealing with a single frequency, the spectral density will be equivalent to the intensity distribution since, in arbitrary units, we have that
\begin{equation}
    I(\mathbf{x}) = \int_{-\infty}^{+\infty} S(\mathbf{x},\omega)\,\mathrm{d}\omega\,.
\end{equation}
Therefore, for a single frequency $\omega_0$, we obtain $I(\mathbf{x})=S(\mathbf{x},\omega_0)$.
The procedure for computing the spectral density through $\mathbf{W}^{(6)}$ is described in Algorithm~\ref{alg:specDen}. 
\begin{algorithm}[t]
\caption{Retrieval of the spectral density array $\mathbf{S}$ for a given CSDA $\mathbf{W}$.}\label{alg:specDen}
\begin{algorithmic}[1]
\Function{specdensity}{${\mathbf{W}},N$}
    \State $\mathbf{S} \gets \textsc{zeros}(N,N)$ \Comment{\textcolor{blue}{Creates a $N\times N$ array with  zeros.}}
    \For{$i=0$ to $N-1$ \textbf{step} 1}
      \For{$j=0$ to $N-1$ \textbf{step} 1}
      \State $\mathbf{S}_{i,j} \gets \mathbf{W}_{i,j;i,j}$
      \EndFor
    \EndFor
     \State \Return  $\mathbf{S}$
\EndFunction	
\end{algorithmic}
\end{algorithm}

To compute the spectral degree of coherence we must use the definition of equation~(\ref{eq:defMu}). However, to represent the spectral degree of coherence, we need a 4D array since it depends on pairs of points $\mathbf{x}_1$ and $\mathbf{x}_2$. For the sake of simplicity, we will represent the spectral degree of coherence array as a 2D array by assuming that vector $\mathbf{x}_1$ is a user-defined point $P_1=(P_{1x},P_{1y})$, which according to equation~(\ref{eq:point}) can be represented as
\begin{equation}
    P_1=(P_{1x},P_{1y})=\Delta x \left(i_P-\frac{N}{2},\frac{N}{2}-j_P\right)\,,
\end{equation}
where $\Delta x$ is the spatial resolution at the plane of the CSDA in analysis, $i_P$ and $j_P$ are constant indexes of the CSDA $\mathbf{W}$, i.e., $i_P,j_P\in\{0,1,...,N-1\}$. The procedure for computing the spectral degree of coherence array from a given CSDA $\mathbf{W}$ is described in the function \textsc{coherence} of Algorithm~\ref{alg:dcoh}.
\begin{algorithm}[t]
\caption{Retrieval of the spectral degree of coherence array $\mathbf{dc}$ for a given CSDA $\mathbf{W}$.}\label{alg:dcoh}
\begin{algorithmic}[1]
\Function{coherence}{${\mathbf{W}},N,i_p,j_p$}
    \State $\mathbf{c} \gets \textsc{zeros}(N,N)$ \Comment{\textcolor{blue}{Creates an $N\times N$ complex array with zeros.}}
    \State $\mathbf{S}_0 \gets \mathbf{W}_{i_P,j_P;i_P,j_P}$ \Comment{\textcolor{blue}{Spectral density at point $P_1$}}
    \For{$i=0$ to $N-1$ \textbf{step} 1}
      \For{$j=0$ to $N-1$ \textbf{step} 1}
    \State $\mathbf{S}' \gets \mathbf{W}_{i,j;i,j}$ \Comment{\textcolor{blue}{Spectral density at point $\mathbf{x_2}$}}
      
      \State $\mathbf{c}_{i,j} \gets \mathbf{W}_{i_P,j_P;i,j}/\left(\sqrt{S}'\sqrt{S_0}'\right)$
      \EndFor
    \EndFor
     \State \Return  $\mathbf{c}$
\EndFunction	
\end{algorithmic}
\end{algorithm}

\section{PyWolf's framework and graphical user interface}
\label{sec:pywolf}

\begin{figure*}
\centering
\includegraphics[width=0.90\textwidth]{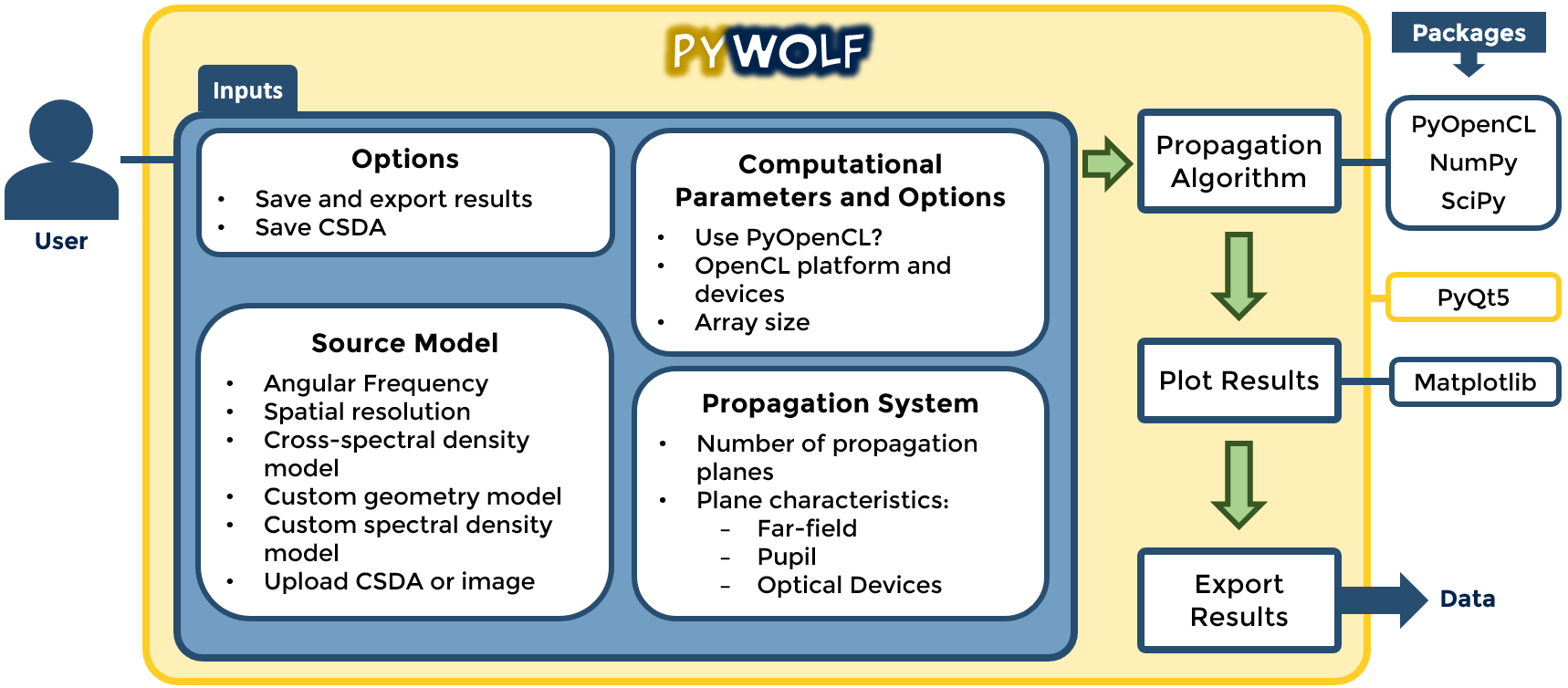}
\caption{\label{fig:GUI} (Color online) Illustration of PyWolf's GUI inputs and main tasks and packages used.}
\end{figure*}

PyWolf contains a graphical user interface (GUI) built with PyQt5, which enables the user to easily perform simulations, plot results, save and load simulation inputs, export results (including CSDAs), and open documentation. Figure~\ref{fig:GUI} illustrates an overview of PyWolf's GUI, summarizing the inputs, the packages used, and main tasks. An advantage of using a user-friendly GUI is that the user can add custom python codes in specific cases  (e.g., source and propagation models) that are recognized by PyWolf's GUI and can be readily selected, along with custom input parameters. We will next describe the main features of PyWolf.

\paragraph{PyOpenCL}
PyWolf detects the available OpenCL platforms and their devices to be used through the PyOpenCL package. The options are then displayed so that the user can choose one. The user can also choose not to use PyOpenCL, in which case sequential computing will be used in the propagation algorithm or in creating the source CSDA  $\mathbf{W}^{(0)}$. 

\paragraph{CSDA size}
The user chooses an input integer value $N$. It will be used to create the CSDA, which is a NumPy complex array of shape $N\times N \times N \times N$.

\paragraph{Source model}
PyWolf enables users to add custom source models by placing Python codes in the source model folder. During initiation, PyWolf scans for source models in that specific folder and adds that option in the GUI. When the user chooses a given source model, the input parameters are requested from the user, For example, if a Python code containing the Gaussian Schell-model source is detected, this option will appear. If it is selected, entries for the effective correlation length $\sigma_\mu$ and the spectral density standard deviation $\sigma_S$ [see equation~(\ref{eq:GSM_source})] will be displayed and the user can fill in their values. Additionally, the user can add two extra options for a given source model: (i) add a custom geometric form, which acts as an optical mask, and (ii) add a custom spectral density. In both cases, the user can add custom-made Python scripts to a specific folder that PyWolf scans and adds as an option in the GUI, along with the respective entries for the input parameters. Lastly, the user can choose to upload a custom CSDA in a NumPy’s npy format, which may have either been exported from a previous simulation or built by the user.

\paragraph{Propagation planes}

After defining the source model, the user must define the propagation system characteristics. First, the user selects the number of propagation planes, located after the
source plane. In the current version, the maximum number is 3. The propagation between two planes obeys equation~\ref{eq:W_fresnel}. For each propagation plane, the user defines the distance $R_i\,(i=1,2,3)$ relative to the previous plane and an option for using the far-field approximation of equation~(\ref{eq:W_farfield}), which decreases the simulation time. Moreover, the user can decide to add a custom pupil (e.g., a circular aperture) and an optical device (e.g., a thin lens). In both cases, the user can add custom-made Python scripts to a specific folder that is detected by PyWolf during initialization.

\paragraph{CSDA and propagation algorithm}
After all required input parameters and options are selected, the user starts the simulation by pressing a specific button. PyWolf starts by creating the class CSDA which contains a $N\times N \times N \times N$ complex array. It then builds the source model according to the user input. The computation time is decreased if the user chooses to use PyOpenCL. Then, PyWolf will propagate the source CSDA to the final observation plane. For instance, if the number of propagation planes is 3, PyWolf implements a loop with three cycles: propagation from the source plane $\mathcal{A}$ to plane $\mathcal{B}$, from plane $\mathcal{B}$ to plane $\mathcal{C}$, from plane $\mathcal{C}$ to plane $\mathcal{D}$. Likewise, if the user decides to use PyOpenCL, the propagation algorithm (Algorithm~\ref{alg:prop}) will be executed at the chosen parallel computing device to decrease the computation time. Note that, in this version, NumPy's FFT is the predefined algorithm to perform the 2D Fourier transforms.

\paragraph{Plot and data export}
After the simulation, the main results are plotted in PyWolf's GUI using the matplotlib repository. For both the source and propagated CSDAs, PyWolf plots three figures: (i) the image (i.e., spectral density for the chosen angular frequency), (ii) the spectral degree of coherence $\mu(P_1,\mathbf{x}_2,\omega_0)$ (3D graph), where $P_1=(P_{1x},P_{1y})$ is a user-defined point, and (iii) the spectral degree of coherence $\mu[P_1,(P_{2x},y_2),\omega_0]$ (2D graph), where $P_1$ is once again a user-defined point and $P_{2x}$ a user-defined coordinate for vector $\mathbf{x}_2$. Finally, the user can export all data, including the source and propagated CSDAs in NumPy's \textit{npy} format.

\begin{table*}[t]
\caption{PyWolf's main input parameters for the 4 simulation examples of Section~\ref{sec:examples}. \label{tab:parameters}}
\begin{centering}
\begin{tabular}{|c|c|c|c|c|}
\hline 
\multirow{2}{*}{Input} & \multicolumn{4}{c|}{Simulation examples}\tabularnewline
\cline{2-5} \cline{3-5} \cline{4-5} \cline{5-5} 
 & 1 & 2 & 3 & 4\tabularnewline
\hline 
\hline 
$N$ & 200 & 200 & 200 & 200\tabularnewline
\hline 
$N_{z}$ & 512 & 512 & - & - \tabularnewline
\hline 
$\Delta\xi$ (m) & $3.51\times10^{-8}$ & $1\times10^{-3}$ & $1\times10^{-6}$ & $1\times10^{-4}$ \tabularnewline
\hline 
$\omega_{0}\,(\mathrm{rad/s)}$ & $1.2566\times10^{19}$ & $3.7699\times10^{15}$ & $2.8560\times10^{15}$ & $2.9920\times10^{15}$\tabularnewline
\hline 
Cross-spectral density model & Coherent & Incoherent & Incoherent & Gaussian Schell-model\tabularnewline
\hline 
\multirow{2}{*}{Cross-spectral density model parameters} & \multirow{2}{*}{-} & \multirow{2}{*}{-} & \multirow{2}{*}{-} & $N_{S}=21$\tabularnewline
\cline{5-5} 
 &  &  &  & $N_{\mu}=41$\tabularnewline
\hline 
Geometry model & Square & Anullar & None & None\tabularnewline
\hline 
\multirow{2}{*}{Geometry model parameters} & \multirow{2}{*}{$N_{a}=57$} & $N_{\mathrm{i}}=5$ & \multirow{2}{*}{-} & \multirow{2}{*}{-}\tabularnewline
\cline{3-3} 
 &  & $N_{\mathrm{o}}=11$ &  & \tabularnewline
\hline 
Geometry from image & False & False & True & False\tabularnewline
\hline 
Number of propagation planes & 1 & 1 & 2 & 1\tabularnewline
\hline 
Distances (m) & $R_{1}=4.244\times10^{-3}$ & $R_{1}=5$ & $R_{1},R_{2}=0.5$ & $R_1=3.1746\,\mathrm{m}$\tabularnewline
\hline 
\end{tabular}
\par\end{centering}
\end{table*}

\section{Simulation examples}
\label{sec:examples}
In what follows, we will present four simulations performed in PyWolf to validate and illustrate our numerical implementation. We will also show how different types of studies can take advantage of PyWolf's PyOpenCL implementation and what are the typical inputs needed. The main inputs for the four simulations are illustrated in Table~\ref{tab:parameters}. Note that the current version of PyWolf is based on the approximation of equation~(\ref{eq:approx_L}) and, therefore, PyWolf's simulations and  the accuracy of its results must take into account its validity.

\subsection{Fresnel propagation of a perfectly coherent squared source}
\label{subsec:Fresnel}

\begin{figure}[t!]
\centering
\includegraphics[width=0.35\textwidth]{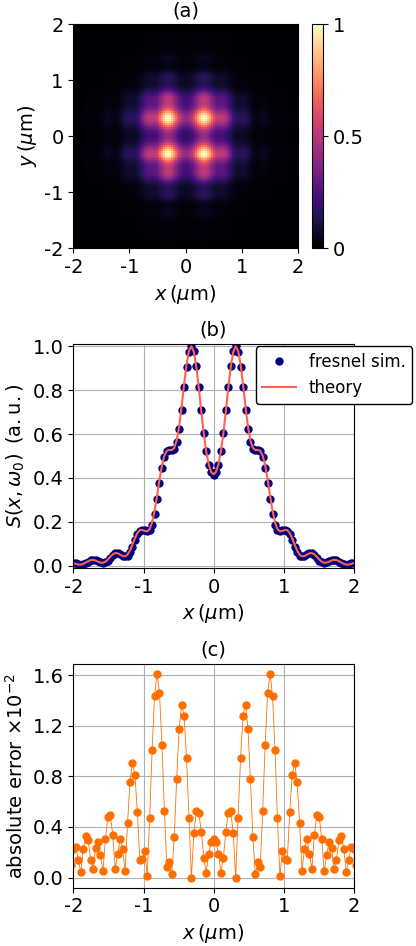}
\caption{\label{fig:Fresnel} (Color online) Fresnel "dark-spot". $\mathbf{(a)}$ Simulated image using PyWolf without the far field approximation (Fresnel). (b) The circular markers are the cross-section of (a) at $y=0\,\mathrm{\mu m}$. The triangular represent the simulation data with the far field approximation. The solid line represents the theoretical values obtained from equation~(\ref{eq:I_square_fresnel}), i.e., without the far field approximation.  (c) Absolute error of the simulation in (b). An independent simulation using equation~(\ref{eq:4error}) was computed and considered as the "true" values for the absolute error. This was done to evaluate the validity of the approximation of equation~(\ref{eq:approx_L}).}
\end{figure}
Let $\rho_1$ be a homogeneous squared source with length $a$, perfectly coherent, quasi-monochromatic with central frequency $\omega_0$ and in plane $\mathcal{A}$. The cross-spectral density function $W_{\rho_1}$ of $\rho_1$ can be written as
\begin{align}
    W_{\rho_1}(\boldsymbol{\xi}_1,\boldsymbol{\xi}_2,\omega_0) = & \,A_1\,\mathrm{rect}\left( \frac{\xi_1}{2a}\right) \mathrm{rect}\left( \frac{\eta_1}{2a}\right)\nonumber \\
     &\times\,\mathrm{rect}\left( \frac{\xi_2}{2a}\right) \mathrm{rect}\left( \frac{\eta_2}{2a}\right)\,,
\end{align}
where $A_1$ is a positive constant. Let $\mathcal{B}$ be an observation plane, distance from plane $\mathcal{A}$ by $R$, where the Fresnel approximation is valid. The intensity $I(\mathbf{x})$ at plane $\mathcal{B}$, using both the Fresnel approximation and that of equation~(\ref{eq:approx_L}), is found to be \cite{Goodman2005}
\begin{align}
    I(\mathbf{x})  = &\frac{1}{4} \left\{\left[C_{\mathrm{F}}(X_{-})-C_{\mathrm{F}}(X_{+})\right]^2+\left[S_{\mathrm{F}}(X_{-})-S_{\mathrm{F}}(X_{+})\right]^2 \right\} \nonumber \\
     & \times \left\{\left[C_{\mathrm{F}}(Y_{-})-C_{\mathrm{F}}(Y_{+})\right]^2+\left[S_{\mathrm{F}}(Y_{-})-S_{\mathrm{F}}(Y_{+})\right]^2 \right\}\,,\label{eq:I_square_fresnel}
\end{align}
where $X$ and $Y$ are modified spatial coordinates defined as
\begin{equation}
    X_{\pm}=\mp\sqrt{\frac{k}{cR}}\left(\frac{a}{2}\pm x\right)\,,
\end{equation}
\begin{equation}
    Y_{\pm}=\mp\sqrt{\frac{k}{cR}}\left(\frac{a}{2}\pm y\right)\,,
\end{equation}
and $C_{\mathrm{F}}(\alpha)$ and $S_{\mathrm{F}}(\alpha)$ are the Fresnel integrals, given by \cite{Goodman1968}
\begin{equation}
    C_{\mathrm{F}}(X) = \int_{0}^{X}\cos \left(\frac{\pi t^2}{2}\right)\,\mathrm{d}t\,,
\end{equation}
\begin{equation}
    S_{\mathrm{F}}(X) = \int_{0}^{X}\sin \left(\frac{\pi t^2}{2}\right)\,\mathrm{d}t\,.
\end{equation}
If the approximation of equation~(\ref{eq:approx_L}) is not valid, the current version of PyWolf will not retrieve valid results. In this case, the intensity $I(\mathbf{x})$ can be calculated through the equivalent of equation~(5) for field propagation in free space  from a primary source~\cite{Yoshimori1995}:
\begin{align}
   I(\mathbf{x}) =\left|\frac{1}{4\pi}\iint Q(\boldsymbol{\xi},\omega_0)\,\frac{\exp\left( \mathrm{i}\,\omega_0 r/c\right)}{r}\,\mathrm{d}\xi \,\mathrm{d}\eta\,\right|^2\,,\label{eq:4error}
\end{align}
where $Q(\boldsymbol{\xi},\omega_0)$ is the source density for a primary source and we used the fact that $I(\mathbf{x})=S(\mathbf{x},\omega_0)$.

For illustrating purposes, we wish to observe the so-called dark-spot in the Fresnel regime at a specific distance from the coherent source. According to the numerical simulations performed by Le Bolloc'h et al.~\cite{le2011x}, the dark spot is located at a distance $R$, given by
\begin{equation}
    R \approx \frac{a^2}{2\pi \lambda_0}\,.\label{eq:R_dark_spot}
\end{equation}
We will next show how to perform the simulation for this propagation system (plane $\mathcal{A}$ to $\mathcal{B}$) using PyWolf. We are mainly interested in comparing the result for the spectral density $S(\mathbf{x},\omega_0)$, at an angular frequency $\omega_0$, with the intensity distribution described by equation~(\ref{eq:I_square_fresnel}).

\begin{figure*}[t!]
\centering
\includegraphics[width=0.95\textwidth]{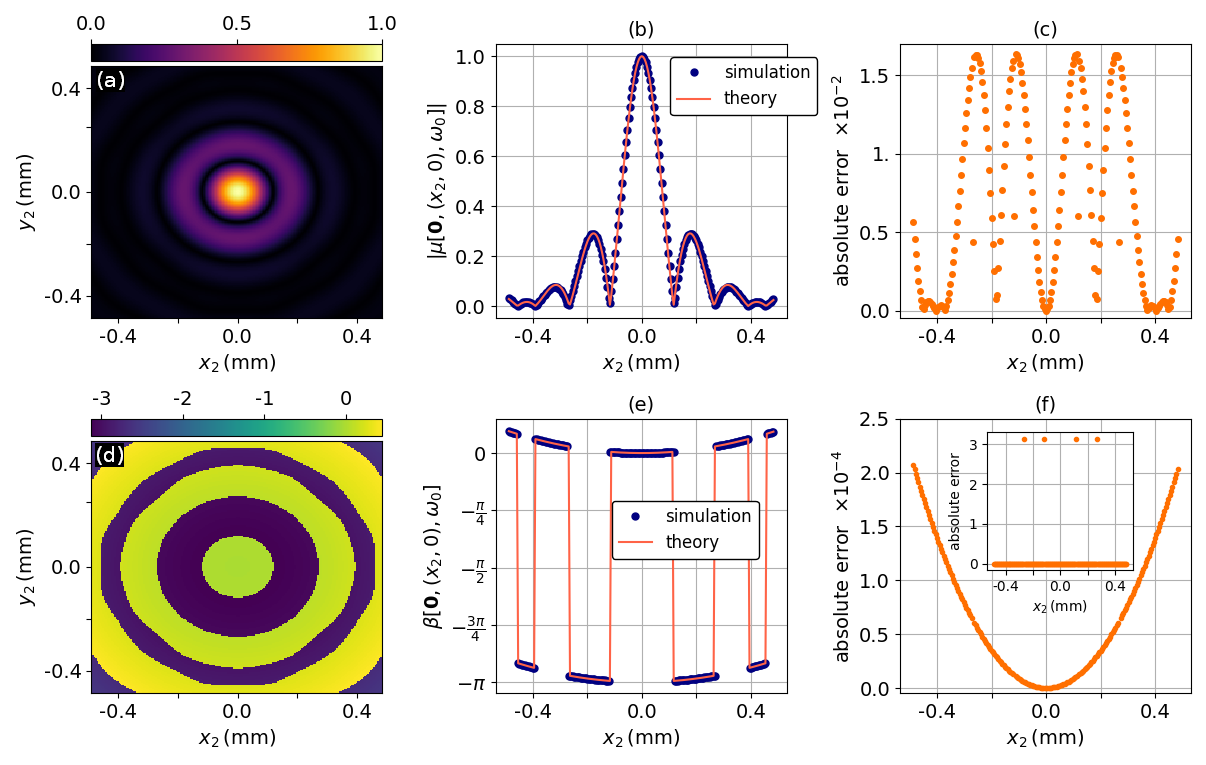}
\caption{\label{fig:Annular} (Color online) Far field spectral degree of coherence of an incoherent annular source. (a) Simulated magnitude of the spectral degree of coherence, $|\mu(\mathbf{0},\mathbf{x}_2,\omega_0)|$ where $\mathbf{0}=(0,0)$ (user-defined). (b) Cross-section of (a) at $y_2=0\,\mathrm{mm}$. The solid lines represent the theoretical values of the magnitude obtained from equation~(\ref{eq:theory_mu_annular}). (c) Absolute error of (b). (d) Phase of the spectral degree of coherence, $\beta(\mathbf{0},\mathbf{x}_2,\omega_0)$. (e) Cross-section of (d) at $y_2=0\,\mathrm{mm}$. The solid lines represent the theoretical values of the phase obtained from equation~(\ref{eq:theory_mu_annular}). (f) Absolute error of (e). The inset shows 4 points which have a larger error. These points occur due to the error in determining the exact position $x$ where the spectral degree of coherence sign switches.}
\end{figure*}

To construct the source CSDA $\mathbf{W}^{(0)}$ with a given spatial resolution step of $\Delta \xi$, we can write the size of the square $a$ as a function of $\Delta \xi$ and the number of elements in each axis, i.e., 
\begin{equation}
    a=N_a \Delta \xi\,,
\end{equation}
where $N_a$ is an adimensional positive integer, defining the number of "pixels" on the square's side, which is an input to PyWolf. The dark spot distance [equation~(\ref{eq:R_dark_spot})] is therefore given by
\begin{equation}
    R = \frac{\left(N_{a} \Delta \xi \right)^2}{2\pi \lambda_0}\,.\label{eq:darkspot}
\end{equation} 
For convenience, we will impose that the resulting minimum spatial sampling (i.e., spatial resolution) $\Delta x$ in plane $\mathcal{B}$ is equal to that of the source plane $\mathcal{A}$, i.e., $\Delta x = \Delta \xi$. Thus, substituting $\Delta x=\Delta \xi$ in equation~(\ref{eq:spatial_resolution}),  substituting equation~(\ref{eq:R_dark_spot}) into~(\ref{eq:spatial_resolution}), and solving for $N_a$, we find that 
\begin{equation}
    N_a = \sqrt{2\pi N}\,.\label{eq:Na}
\end{equation}
Thus, the values $R$ and $N_a$ are now functions of $N$ (or $N_z$ when a zero-padding is used), $\lambda$, and $\Delta \xi$.

\begin{figure*}
\centering
\includegraphics[width=0.95\textwidth]{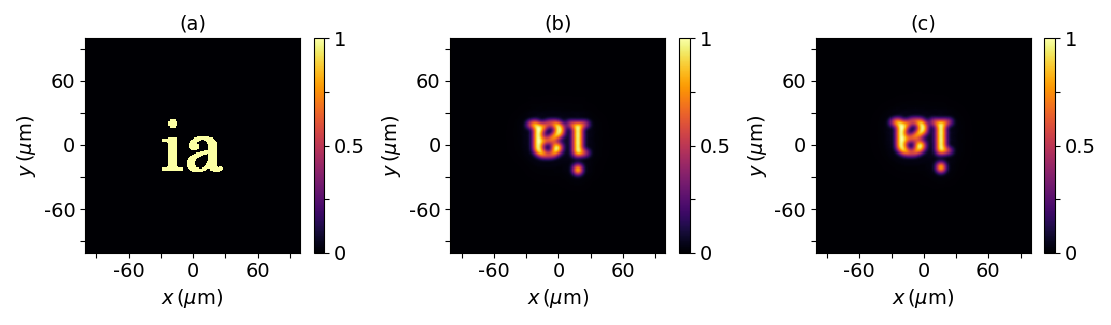}
\caption{\label{fig:ia} (Color online) Imaging partially coherent light. (a) Spectral density $S(\boldsymbol{\xi},\omega_0)$ at the source plane $\mathcal{A}$ using the CSDA $\mathbf{W}^{(0)}$. (b) Simulated spectral density $S(\boldsymbol{\xi}',\omega_0)$ at the image plane $\mathcal{C}$ using the CSDA $\mathbf{W}^{(6)}$. (c) Simulated image using equation~(\ref{eq:ICincoherent}) and the image in (a).}
\end{figure*}

To perform the simulation in PyWolf, we start by defining the CSDA size parameter $N=200$. This choice is driven by the amount of system memory. Additionally, we will use a zero padding procedure in the FFT of $N_\mathrm{z}=512$ and, therefore, equation~(\ref{eq:Na}) gives $N_a\approx57$. For the spatial resolution we set $\Delta\xi=0.0351\,\mathrm{\mu m}$ so that $a=2\,\mathrm{\mu m}$. For the central frequency we set $\omega_0=1.2566\times10^{19}\,\mathrm{rad/s}$ (in free space, $\lambda_0\approx1.5\,\AA$). The distance between the source and observation planes is set as the dark spot distance of equation~(\ref{eq:darkspot}), which gives $R_1=4.244\,\mathrm{mm}$. Notice that, in this case, the paraxial approximation of equation~(\ref{eq:approx_L})~is valid. The Fresnel number $N_F$ in this case (see, for example, Chapter 4.4 of reference~\cite{Goodman2005}) is:
\begin{eqnarray}
    N_F = \frac{a^2}{\lambda_0\, R}= \frac{\left(N_a\Delta\,\xi\right)^2}{\lambda_0\, R}\approx 6.29\,,
\end{eqnarray}
and, therefore, the Fresnel and far field approximations will have significant differences. Then, we chose perfectly coherent light for the cross-spectral density model. Table~\ref{tab:parameters} summarizes the input values in PyWolf for this simulation (Simulation example 1). Results for the spectral density are shown in Fig.~\ref{fig:Fresnel} and compared with the theoretical ones which are calculated using equation~(\ref{eq:I_square_fresnel}). The absolute error is computed with an independent simulation using equation~(\ref{eq:4error}).

\subsection{Far-field propagation of an incoherent annular source}
\label{sec:fre}

Let us now consider the case of a homogeneous, perfectly incoherent, and quasi-monochromatic annular source $\rho_2$ with inner and outer radius $a_{\text{i}}$ and $a_{\text{o}}$, respectively, and central angular frequency $\omega_0$. We can write the cross-spectral density function of $\rho_2$ at the source plane $\mathcal{A}$ as
\begin{align}
    W_{\rho_2}(\boldsymbol{\xi}_1,\boldsymbol{\xi}_2,\omega_0)  = &\, A_2 \left[ \mathrm{circ}\left(\frac{\boldsymbol{\xi}_1}{a_\mathrm{o}}\right) \mathrm{circ}\left(\frac{\boldsymbol{\xi}_2}{a_\mathrm{o}}\right)-\mathrm{circ}\left(\frac{\boldsymbol{\xi}_1}{a_\mathrm{i}}\right)\mathrm{circ}\left(\frac{\boldsymbol{\xi}_1}{a_\mathrm{i}}\right) \right] \nonumber \\
    & \times\,\delta(|\boldsymbol{\xi}_2-\boldsymbol{\xi}_1|,\omega_0)  \,, 
\end{align}
where $A_2$ is a positive constant and $\mathrm{circ}(\boldsymbol{\xi})$ is the circle function defined as
\begin{equation}
    \mathrm{circ}(\boldsymbol{\xi})=\begin{cases}
    1 & ,\,|\mathrm{\boldsymbol{\xi}}|<1\\
    \frac{1}{2} & ,\,|\mathrm{\boldsymbol{\xi}}|=1\\
    0 & ,\,|\mathrm{\boldsymbol{\xi}}|>1
    \end{cases}\,.
\end{equation}
Let $\mathcal{B}$ be the observation plane located in the far-field at a distance $R_1$ from the source plane $\mathcal{A}$. For the sake of brevity, let us define $b=|\mathbf{x}_2-\mathbf{x}_1|$ and $b'=\mathbf{x}^2_2-\mathbf{x}^2_1$. The spectral degree of coherence at $\mathcal{B}$ can be calculated using the van Cittert-Zernike theorem~\cite{MandelWolf1995}:
\begin{align}
    \mu(b,b',\omega_0) =& \left\{ \left[\frac{2\mathrm{J}_1\left(k a_\mathrm{o}b/R_1\right)}{k a_\mathrm{o}b/R_1} \right] -\frac{a_\mathrm{i}}{a_\mathrm{o}} \left[\frac{2\mathrm{J}_1\left(k a_\mathrm{i}b/R_1\right)}{k a_\mathrm{i}b/R_1} \right] \right\} \nonumber \\
    & \times \exp \left(-\mathrm{i}\frac{\omega_0\,b'}{c\,R} \right) \,,\label{eq:theory_mu_annular}
\end{align}
where $\mathrm{J}_1$ is the first-order Bessel function of the first kind. We seek to simulate the propagation of the cross-spectral density function of source $\rho_2$, extract the spectral degree of coherence and compare it with the theoretical values of equation~(\ref{eq:theory_mu_annular}).

To build the source geometric model of the source, we defined two adimensional parameters for the inner and outer radius, $N_\mathrm{i}$ and $N_\mathrm{o}$, respectively, such that
\begin{equation}
    a_\mathrm{i} = N_\mathrm{i}\,\Delta \xi\,,
\end{equation}
\begin{equation}
    a_\mathrm{o} = N_\mathrm{o}\,\Delta \xi\,.
\end{equation}
To perform this simulation, we chose the array size parameter $N=200$ and did a zero padding for the FFT with a total size of $N_{\mathrm{z}}=512$, similar to what was used in the examples of Section~\ref{subsec:Fresnel}. For the central wavelength, we used $\omega_0=3.7699\times10^{15}\,\mathrm{rad/s}$ (yellow color), while for the spatial resolution we chose $\Delta \xi=1\times10^{-3}\,\mathrm{m}$. For the adimensional parameters of the source geometric model, we chose $N_\mathrm{i}=5$ and $N_\mathrm{o}=11$. The distance between the source and observation planes is set to $R_1=5\,\mathrm{m}$, which is sufficient for a valid far-field approximation. Table~\ref{tab:parameters} summarizes the input values (Simulation example 2). Results for the magnitude and phase of the simulated spectral degree of coherence are illustrated in Fig.~\ref{fig:Annular} as well as the theoretical values for comparison using equation~(\ref{eq:theory_mu_annular}). Notice that the exponential term in equation~(\ref{eq:theory_mu_annular}) adds an additional phase term to the spectral degree of coherence.
\begin{figure*}[t!]
\centering
\includegraphics[width=0.85\textwidth]{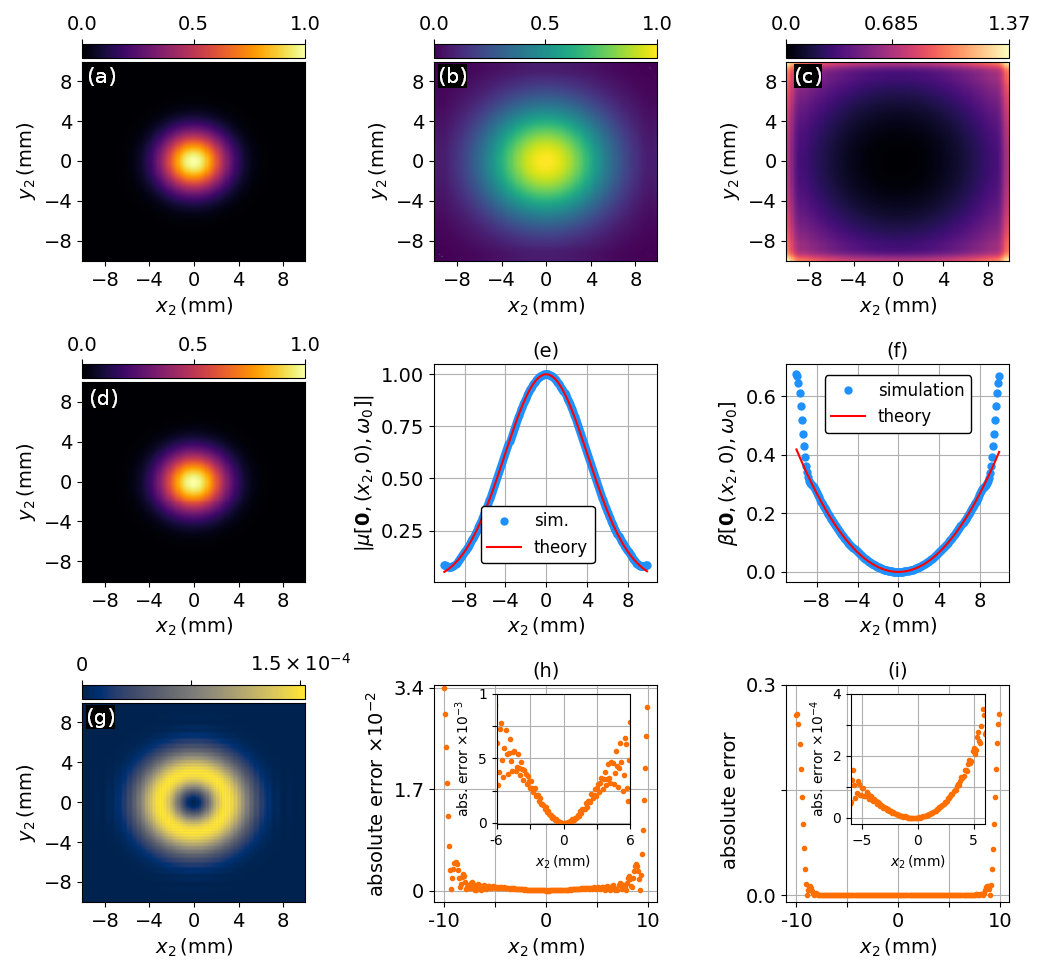}
\caption{\label{fig:beam} (Color online) Gaussian Schell-model beam. (a) Spectral density $S(\boldsymbol{\xi},\omega_0)$ at the source plane $\mathcal{A}$ using the CSDA $\mathbf{W}^{(0)}$. (b) Simulated magnitude of the spectral degree of coherence, $|\mu(\mathbf{0},\mathbf{x}_2,\omega_0)|$ for $\mathbf{0}=(0,0)$. (c) Simulated phase of the spectral degree of coherence, $\beta(\mathbf{0},\mathbf{x}_2,\omega_0)$. (d) Simulated spectral density $S(\boldsymbol{\xi}',\omega_0)$ at the observation plane $\mathcal{B}$ using the CSDA $\mathbf{W}^{(6)}$. (e) Cross-section of (b) at $y_2=0\,\mathrm{mm}$ with the theoretical values using equation~(\ref{eq:Wbeam}). (i) Cross-section of (c) at $y_2=0\,\mathrm{mm}$ with the theoretical values using equation~(\ref{eq:Wbeam}). (g) Absolute error of the simulated spectral density represented in (d) calculated using the expected values. (h), (i) absolute errors for the magnitude and phase of the simulated spectral degree of coherence, respectively. The insets show the absolute error in a smaller interval of $x$, namely, $\left[-6,6\right]$ in (h) and $\left[-5,5\right]$ in (i).}
\end{figure*}

\subsection{Aperture effects in an imaging system with partially coherent light}
\label{sec:imaging}

We will now use PyWolf to simulate the propagation of light in thin lens imaging, as illustrated in Fig.~\ref{fig:imaging}. The goal is to validate PyWolf's simulation results in the case where partially coherent light impinging on an aperture, such as that of a lens, produces changes in the image of an object. This type of problem can be found, for instance, in transfer functions of astronomical telescopes or microscopies, although, in this case, we tackle it with the cross-spectral density function.

Let $\rho_3$ be a perfectly incoherent, homogeneous, and quasi-mono\-chromatic light source with central frequency $\omega_0$, located in plane $\mathcal{A}$. Suppose that a mask is placed at the source plane so that it changes its shape. The cross-spectral density can be written as
\begin{equation}
    W_{\mathcal{A}}(\boldsymbol{\xi}_1,\boldsymbol{\xi}_2,\omega_0)=Y(\boldsymbol{\xi}_1)^{*}\,W_{\rho_3}(\boldsymbol{\xi}_1,\boldsymbol{\xi}_2,\omega_0)\,Y(\boldsymbol{\xi}_2)\,,
\end{equation}
where $Y(\boldsymbol{\xi})$ is a function that defines the geometry of the source with counter domain $\{0,1\}$ and $W_{\rho_3}(\boldsymbol{\xi}_1,\boldsymbol{\xi}_2,\omega_0)$ is the cross-spectral density for a perfectly incoherent source, given by
\begin{equation}
    W_{\rho_3}(\boldsymbol{\xi}_1,\boldsymbol{\xi}_2,\omega_0)=A_3\,\delta(\boldsymbol{\xi}_1-\boldsymbol{\xi}_2,\omega_0)\,,
\end{equation}
where $A_3$ is a positive constant. In this case, the source follows the letters "ia" [see Fig.~\ref{fig:ia}(a)]. 

A positive and circular lens, L, with focal length $f$ and radius $a_\mathrm{L}$, is placed at a distance $R_1$ from the source, in plane $\mathcal{B}$. Similar to previous examples, we define an adimensional parameter $N_{\mathrm{L}}$ for the aperture radius $a_{\mathrm{L}}$ such that
\begin{equation}
    a_{\mathrm{L}} = N_{\mathrm{L}}\,\Delta\xi \,.
\end{equation}
Light from plane $\mathcal{A}$ reaching the lens plane, $\mathcal{B}$, will be partially coherent and modeled by the van Cittert-Zernike theorem. An image will be formed at plane $\mathcal{C}$, distanced by $R_2$. For an ideal imaging system, $R_1$ and $R_2$ satisfy equation~(\ref{eq:imaging}). The quality of the image of an incoherent object will depend on the spatial coherence area of light reaching plane $\mathcal{B}$ and on the aperture size of the lens. On the other hand, the spatial coherence area of light reaching the lens will depend on the wavelength of light and the distance $R_1$. In terms of coherence theory, the cross-spectral density of $W_\mathcal{C}(\boldsymbol{\xi}_1',\boldsymbol{\xi}_2',\omega_0)$ at the image plane will be given by equation~(\ref{eq:WC_imaging}) and the spectral density (or the intensity) is directly extracted from $W_\mathcal{C}(\boldsymbol{\xi}_1',\boldsymbol{\xi}_2',\omega_0)$. However, since the source is perfectly incoherent, instead of using coherence theory to describe the image in plane $\mathcal{C}$, we can use a well-known result from Fourier optics which states that for incoherent objects, the intensity $I_{\mathcal{C}}(\xi',\eta')$ of the image is given by (see, for example, Section 6.3 of reference~\cite{Goodman2005}):
\begin{equation}
    I_{\mathcal{C}}(\xi',\eta')=\iintop_{-\infty}^{+\infty}p(\xi-\xi',\eta-\eta')\, I_{A}(\xi,\eta)\,\mathrm{d}\xi\,\mathrm{d}\eta\,\label{eq:ICincoherent}\,,
\end{equation}
where $p(x,y)$ is the impulse response given by:
\begin{equation}
    p(x,y)=\left|\frac{\lambda_{0}R_{2}}{\pi a\sqrt{x^{2}+y^{2}}}\mathrm{J}_{1}\left(\frac{2\pi a\sqrt{x^{2}+y^{2}}}{\lambda_{0}R_{2}}\right)\right|^{2}\,.
\end{equation}
To validate our results, we will compare our results with that of equation~(\ref{eq:ICincoherent}).

To perform this simulation, we set the size parameter $N=200$, without zero padding. Then, we choose the central angular frequency to $\omega_0=2.8560\times10^{15}\,\mathrm{rad/s}$ (red color) and a spatial resolution of $\Delta\xi=1\,\mathrm{\mu m}$. To define the source geometry, we uploaded a binary image of size $200\times200$ containing the letters "ia". Then, we selected two propagation planes, $\mathcal{B}$ (lens plane) and $\mathcal{C}$ (image plane). At plane $\mathcal{B}$, we set the circular aperture parameter to $N_{\mathrm{L}}=11$. The spatial resolution $\Delta x$ at plane $\mathcal{B}$ is calculated using equation~(\ref{eq:spatial_resolution}), which gives $\Delta x = 1.65\,\mathrm{mm}$. Therefore, the lens aperture radius is $a_{\mathrm{L}}=1.1815\,\mathrm{cm}$. For the distances between planes $\mathcal{A}-\mathcal{B}$ and $\mathcal{B}-\mathcal{C}$ we chose $R_1=0.5\,\mathrm{m}$ and $R_2=0.5\,\mathrm{m}$, respectively. To comply with equation~(\ref{eq:imaging}), the focal length of the lens was set to $f=0.25\,\mathrm{m}$. With these values, the lateral magnification is -1 and the spatial resolution in plane $\mathcal{C}$ is the same as that of the source plane, i.e., $\Delta\xi'=\Delta\xi=1\,\mathrm{\mu m}$. Table~\ref{tab:parameters} summarizes the input values in PyWolf for this simulation (Simulation example 3). The source spectral density is illustrated in Fig.~\ref{fig:ia}(a) and the results for the simulated spectral density in the image plane are shown in Fig.~\ref{fig:ia}(b). To compare the results obtained by PyWolf, Fig.~\ref{fig:ia}(c) shows the result for the same simulation using equation~(\ref{eq:ICincoherent}), i.e., using Fourier optics.
\begin{table*}[t!]
\begin{centering}
\protect\caption{Characteristics of each computer system (CS) used for the performance analysis.\label{tab:performance}}
\begin{tabular}{|c|c|c|c|c|c|}
\hline 
\multirow{2}{*}{CS} & \multirow{2}{*}{CPU} & \multirow{2}{*}{RAM} & \multirow{2}{*}{GPU} & \multicolumn{2}{c|}{OpenCL}\tabularnewline
\cline{5-6} 
 &  &  &  & Platform & Device\tabularnewline
\hline 
\hline 
\multirow{2}{*}{1} & \multirow{2}{*}{Intel i7-6950X} & \multirow{2}{*}{128 GB} & \multirow{2}{*}{Nvidia GTX 1050 (4 GB)} & Nvidia CUDA & GTX 1050\tabularnewline
\cline{5-6} 
 &  &  &  & Intel OpenCL & i7-6950X\tabularnewline
\hline 
\multirow{2}{*}{2} & \multirow{2}{*}{Intel i5-4690} & \multirow{2}{*}{32 GB} & \multirow{2}{*}{AMD Radeon R9 380 (4 GB)} & AMD APP & R9 380\tabularnewline
\cline{5-6} 
 &  &  &  & AMD APP & i5-4690\tabularnewline
\hline 
\multirow{2}{*}{3} & \multirow{2}{*}{Intel i5-3230M} & \multirow{2}{*}{8 GB} & \multirow{2}{*}{Nvidia GT 740M (2 GB)} & Nvidia CUDA & 740M\tabularnewline
\cline{5-6} 
 &  &  &  & Intel OpenCL & i5-3230M\tabularnewline
\hline 
\end{tabular}
\par\end{centering}
\end{table*}

\subsection{Beam condition for a Gaussian Schell-model source}
\label{sec:beam_condition}

Let $\rho_4$ be a planar, secondary, quasi-monochromatic Gaussian Schell-model source with central frequency $\omega_0$ and contained in plane $\mathcal{A}$. Its cross-spectral density is given by~\cite{Wolf2007}
\begin{align}
    W_{\rho_4}(\boldsymbol{\xi}_1,\boldsymbol{\xi}_2,\omega_0)=A_4 \exp\left[\frac{\left(\boldsymbol{\xi}_1^2+\boldsymbol{\xi}_2^2\right)}{4\sigma_S^2} \right] \exp\left[\frac{\left|\boldsymbol{\xi}_1-\boldsymbol{\xi}_2\right|^2}{2\sigma_\mu^2} \right]\,,
\end{align}
where $\sigma_mu$ and $\sigma_S$ are the effective correlation length and standard deviation of the spectral density, respectively. It can be shown (see Chapter 5 of reference~\cite{MandelWolf1995} that this source model can generate a beam if the following necessary and sufficient condition is fulfilled:
\begin{equation}
    \frac{1}{4\sigma_S^2}+\frac{1}{\sigma_\mu^2} \ll \frac{2\pi^2}{\lambda^2}\,.
\end{equation}
If such is the case, the cross-spectral density at a given observation plane $\beta$ can be shown to be~\cite{MandelWolf1995}
\begin{align}
W(\mathbf{x}_{1},\mathbf{x}_{2},\omega) = & \exp\left[-\frac{1}{4\left(\psi\psi^{*}-\beta^{2}\right)}\left(\psi\mathbf{x}_{1}^{2}+\psi^{*}\mathbf{x}_{2}^{2}-2\beta\mathbf{x}_{1}\cdot\mathbf{x}_{2}\right)\right]\nonumber \\
  & \times \frac{A^{2}(\omega)}{16\left(a^{2}-b^{2}\right)\left(\psi\psi^{*}-\beta^{2}\right)}\,,\label{eq:Wbeam}
\end{align}
where $a$, $b$, $\beta$ and $\psi$ are parameters defined as
\begin{equation}
    a = \frac{1}{4\sigma_{S}^{2}}+\frac{1}{2\sigma_{\mu}^{2}}\,,
\end{equation}
\begin{equation}
    b = \frac{1}{2\sigma_{\mu}^{2}}\,,
\end{equation}
\begin{equation}
    \beta = \frac{b}{4\left(a^{2}+b^{2}\right)}\,,
\end{equation}
\begin{equation}
    \psi = \alpha-\mathrm{i}\frac{cR}{2\omega}\,.
\end{equation}
The spectral density and spectral degree of coherence can be readily extracted from equation~(\ref{eq:Wbeam}) by using definitions in equations~(\ref{eq:defS}) and~(\ref{eq:defMu}). We will therefore use PyWolf to simulate the propagation of this source model and compare the numerical results with that of equation~(\ref{eq:Wbeam}).

To perform this simulation in PyWolf, we set the size parameter to $N=200$, without zero padding. The angular frequency and spatial resolution chosen were $\omega_0=2.9920\times10^{15}\,\mathrm{rad/s}$ ($\lambda_0=630\,\mathrm{nm}$) and $\Delta\xi=0.1\,\mathrm{mm}$, respectively. We then chose the Gaussian Schell-model source option in PyWolf. To construct the CSDA for this model, we need two adimensional parameters, $N_S$ and $N_u$, defined as
\begin{equation}
    N_S=\frac{\sigma_S}{\Delta \xi}\,,
\end{equation}
\begin{equation}
    N_\mu=\frac{\sigma_\mu}{\Delta \xi}\,.
\end{equation}
For these parameters, we chose $N_S=21$ and $N_u=41$. The distance between the source plane and observation plane was set to $R_1=3.1746\,\mathrm{m}$ so that $\Delta\xi =\Delta x$. The far-field approximation is not used. Table~\ref{tab:parameters} summarizes the input values in PyWolf for this simulation (Simulation example 4) and the simulation results are represented in Fig.~\ref{fig:beam}.

\section{Performance}
\label{sec:performance}

\begin{figure*}[t!]
\centering
\includegraphics[width=0.85\textwidth]{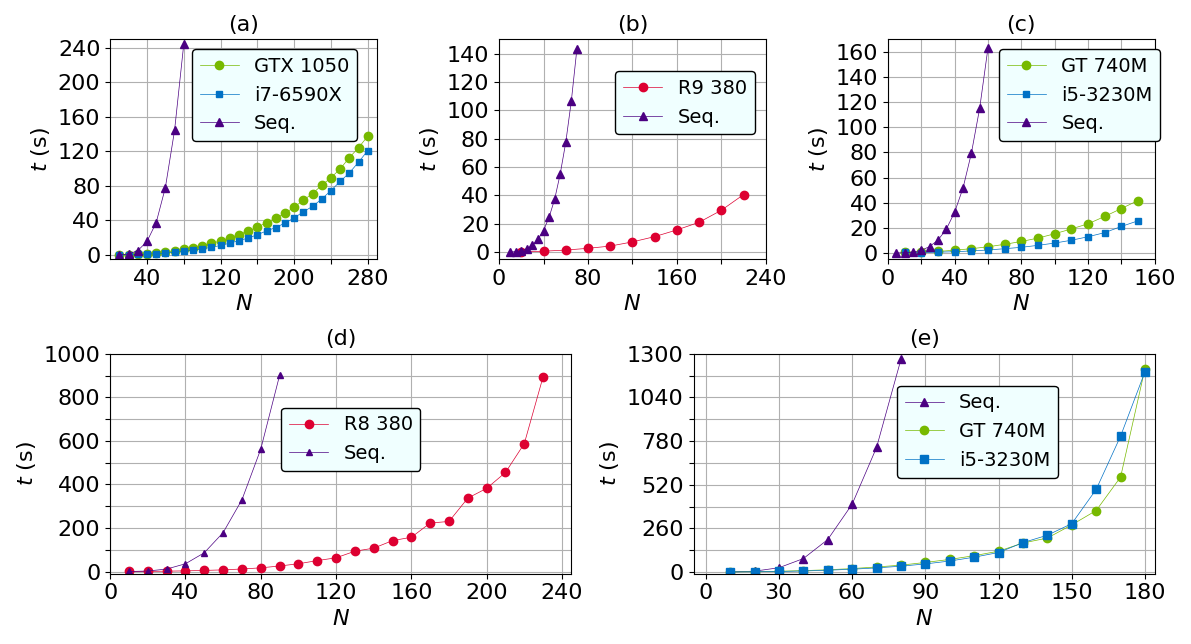}
\caption{\label{fig:performance} (Color online) Computation time of the CSDA $\mathbf{W}^{(0)}$ versus the matrix size parameter size $N$ (a-c) for three computer systems (see Table~\ref{tab:parameters}) and total computation time of the simulation example 2 of Section~\ref{sec:fre} as a function of the matrix size parameter $N$ (d-e). (a) Computer system 1. The triangular markers represent the computation without using PyOpenCL, while the other markers use PyOpenCL with a different platform and device. (b) Computer system 2. The triangular markers represent the computation without using PyOpenCL, while the other marker uses PyOpenCL with the graphics card AMD Radeon 380. (c) Computer system 3. The triangular markers represent the computation without using PyOpenCL, while the other two types of markers represent the use of PyOpenCL with two different platforms and devices. (d) Computer system 2. The triangular markers represent the computation without using PyOpenCL, while the other marker uses PyOpenCL with the graphics card AMD Radeon 380. (e) Computer system 3. The triangular markers represent the computation without using PyOpenCL, while the other two types of markers represent the use of PyOpenCL with two different  devices.}
\end{figure*}
The first significant time-consuming process is the creation of an empty array of size $N^4$, which depends on the available RAM. Python's NumPy package is used for such a task. Then, PyWolf builds the CDSA $\mathbf{W}^{(0)}$ according to the user-defined model. In general, PyWolf will have to compute $N^4$ values, which is a heavy time-consuming task. If the user selects the option of using PyOpenCL, PyWolf will use the selected Open\-CL's platform and device to parallelize this task.

To evaluate the advantage of using parallel computing by means of Py\-OpenCL, we performed the computation of the CSDA $\mathbf{W}^{(0)}$ for a Gaussian Schell-model source to compare the time computation when using PyOpenCL and when using only sequential computing (see Algorithm~\ref{alg:GSM} in \ref{sec:GSMS_example}). We chose three computers with the same operating system, Windows 10, to execute PyWolf and perform the same task, namely, the creation of a CSDA of a Gaussian Schell-model source. Table~\ref{tab:performance} summarizes computers’ specifications. Besides the difference in the central processing units (CPUs) and the graphical processing units (GPUs), the random access memory (RAM) is also different, which may limit the maximum value for the input size parameter $N$. For each computer, we varied the size parameter $N$ to analyze how the computation time evolves. Figures~(\ref{fig:performance})(a-c) shows results for the computation of the CSDA $\mathbf{W}^{(0)}$ of a Gaussian Schell-model source. Clearly, using PyOpenCL is a significant advantage in terms of computation time. 

For the purpose of evaluating the advantage of using PyWolf's PyOpenCL implementation in a full simulation for the propagation of partially coherent light, i.e. from the creation of the source CSDA $\mathbf{W}^{(0)}$ to the final computation of the CSDA $\mathbf{W}^{(6)}$, we performed simulation example 2 (see Section~\ref{sec:beam_condition}), with and without using PyOpenCL, in two different computer systems. Figures~\ref{fig:performance}(e-f) shows the computation time results for the computation of the final CSDA $\mathbf{W}^{(0)}$ as a function of the array size parameter $N$. In terms of computation time, the advantage of using PyOpenCL is easily observed as $N$ increases.

\section{Conclusion}
\label{sec:conclusion}

We introduced PyWolf, an open-source Python software that performs simulations of propagation of partially coherent light in diffraction and imaging. This program takes advantage of the open-source toolkit PyOpenCL to perform parallel computation in order to reduce the computation time. It allows the user to easily add new options in terms of source or propagation models. PyWolf has a graphical user interface built with PyQt5 which not only enables the user to choose simulation optics easily but also recognizes new custom codes added to PyWolf, such as source models, geometries, spectral densities, and pupil functions. After each simulation, the user can retrieve optical quantities such as the spectral degree of coherence and spectral density. This allows users, for instance, to evaluate the impact of each source model in a given optical system. For a given simulation, the numerical error will stem mainly from the spatial sampling used for a given source model, which will affect the outcome of the Fourier transform algorithm - which in this first version of PyWolf is the NumPy’s FFT algorithm - and, therefore, any simulation must take into account the errors introduced by the sampling.

Future improvements of PyWolf can include the use of different transform algorithms, including other approaches to propagation from radially symmetric source models (e.g., the Hankel transform). In the current version, PyWolf only simulates the propagation of the cross-spectral density for a given frequency. Future versions will also include the option of using polychromatic light. Simulations of correlation-induced spectral changes can also be added, as done in reference~\cite{Magalhaes2018}. Computationally, the main bottleneck of PyWolf is the system memory usage, i.e., RAM. As the size of the CSDA increases, the memory increases by a factor of $N^4$. PyWolf is intended to be a community-oriented open-source software, on behalf of the optical coherence community and we welcome and will acknowledge contributions for other computational methods, algorithms, and custom models.

\section*{Acknowledgment}
Tiago E. C. Magalh{\~a}es acknowledges the support from Fun\-da\-\c{c}{\~a}o para a Ci\-{\^e}n\-cia e a Tec\-no\-lo\-gia (FCT, Portugal) through the grant PD/BD/105952/2015, under the FCT PD Program \\ PhD::SPA\-CE (PD/00040/2012). This work was supported by \\ FCT/MCTES through national funds (PIDDAC) by this grant \\ UID/FIS/04434/2019. The authors are grateful to the referees for their suggestions.

\appendix

\section{Example of a CSDA $\mathbf{W}^{(0)}$ for a Gaussian Schell-model source in a circular aperture}

\label{sec:GSMS_example}

\begin{algorithm}[b!]
\caption{Creation of the source CSDA with a circular geometry using sequential or parallel computing.}\label{alg:circle}
\begin{algorithmic}[1]
\Function{Circle}{${\mathbf{W}},N, N_a, \text{usePyOpenCL}$}
    \For{$i_1=0$ to $N-1$ \textbf{step} 1}
      \For{$j_1=0$ to $N-1$ \textbf{step} 1}
        \State$r_1\gets\sqrt{\left(i_1-\frac{N}{2}\right)^2+\left(\frac{N}{2}-j_1 \right)^2} 
        $
\If{$r_1\leq N_a$}
\State $\mathbf{W}_{i_1,j_1}\gets \mathrm{ones}(N,N)$
\Else

\If{$\text{usePyOpenCL}=\mathrm{True}$}
\State \For{ $i_2,j_2=0$ to $N-1$ \textbf{step} 1} \textbf{in parallel}
\Indent
    \State$r_2\gets \sqrt{\left(i_2-\frac{N}{2}\right)^2+\left(\frac{N}{2}-j_2 \right)^2 }$
    \If{$r_2\leq a$}
\State $\mathbf{W}_{i_1,j_1;i_2,j_2}\gets\mathrm{ones}(N,N)$
    \EndIf
\EndIndent
    \EndFor
    \Else
      \For{$i_2=0$ to $N-1$ \textbf{step} 1}
      \For{$j_2=0$ to $N-1$ \textbf{step} 1}
    
    \State$r_2\gets \sqrt{\left(i_2-\frac{N}{2}\right)^2+\left(\frac{N}{2}-j_2 \right)^2 }$
    \If{$r_2\leq N_a$}
\State $\mathbf{W}_{i_1,j_1;i_2,j_2}\gets\mathrm{ones}(N,N)$
\EndIf
         \EndFor
    \EndFor
    \EndIf
    \EndIf
      \EndFor
    \EndFor
     \State \Return  $\mathbf{W}$
\EndFunction	
\end{algorithmic}
\end{algorithm}
\begin{algorithm}[t]
\caption{Creation of a Gaussian Schell-model (GMS) source using sequential or parallel computation.}\label{alg:GSM}
\begin{algorithmic}[1]
\Function{GSMsource}{${\mathbf{W}},N, N_S, N_{\mu},\text{usePyOpenCL}$}
    \For{$i_1=0$ to $N-1$ \textbf{step} 1}
      \For{$j_1=0$ to $N-1$ \textbf{step} 1}
      \State $r_1\gets(i_1-N/2)^2-(N/2-j_1)^2$
\If{$\text{usePyOpenCL}=\mathrm{True}$}
\State \For{ $i_2,j_2=0$ to $N-1$ \textbf{step} 1} \textbf{in parallel}
\Indent
    \State $r\gets(i_1-i_2)^2-(j_2-j_1)^2$ 
    \State $r_2\gets(i_2-N/2)^2-(N/2-j_2)^2$ 
    \State $\mathbf{W}_{i_1,j_1;i_2,j_2}' \leftarrow \mathbf{W}_{i_1,j_1;i_2,j_2}\,\exp \left[-\frac{r}{2N_{\mu}^2} \right] $ \phantom . \phantom .  $\times \exp \left[-\frac{r_1+r_2}{2N_{S}^2} \right]$
\EndIndent
    \EndFor
    \Else
      \For{$i_2=0$ to $N-1$ \textbf{step} 1}
      \For{$j_2=0$ to $N-1$ \textbf{step} 1}
    \State $r\gets(i_1-i_2)^2-(j_2-j_1)^2$ 
    \State $r_2\gets(i_2-N/2)^2-(N/2-j_2)^2$ 
    \State $\mathbf{W}_{i_1,j_1;i_2,j_2}' \leftarrow \mathbf{W}_{i_1,j_1;i_2,j_2}$ \phantom a \phantom a \phantom a \phantom a \phantom a \phantom a \phantom .$\times \exp \left[-\frac{r}{2N_{\mu}^2} \right]  \exp \left[-\frac{r_1+r_2}{2N_{S}^2} \right]$
         \EndFor
    \EndFor
    \EndIf
      \EndFor
    \EndFor
     \State \Return  $\mathbf{W}'$
\EndFunction	
\end{algorithmic}
\end{algorithm}

Suppose we want to create a Gaussian Schell-model source with angular frequency $\omega_0$ bounded by a circle of radius $a$. This could represent, for instance, a quasi-monochromatic laser beam passing through a circular aperture of radius $a$. 

Let $W_{\text{GS}}\left(\boldsymbol{\xi}_1,\boldsymbol{\xi}_2,\omega_0\right)$ be the cross-spectral density of a Gaussian Schell-model source given by equation~(\ref{eq:GSM_source})  and $Y_{\text{c}}(\boldsymbol{\xi})$ a circular aperture function defined as
\begin{equation}
    Y_{\text{c}}(\boldsymbol{\xi})=\begin{cases}
1 & \text{if}\,|\boldsymbol{\xi}|\leq a\\
0 & \text{otherwise}
\end{cases} \label{eq:circle}\,,
\end{equation} 

The cross-spectral density at plane $\mathcal{A}$ will be the product of $W_{\text{GS}}(\boldsymbol{\xi}_1,\boldsymbol{\xi}_2,\omega_0)$ with $Y_{\text{c}}(\boldsymbol{\xi}_1)Y_{\text{c}}(\boldsymbol{\xi}_2)$. To build the source CSDA $\mathbf{W}^{(0)}$, we create a complex $N\times N \times N \times N$ array full of zeros (i.e., $0+0\mathrm{i}$). We then apply the desired geometric shape. This task can be parallelized using PyOpenCL. Algorithm~\ref{alg:circle}  illustrates the procedure for both sequential and parallel computation. It can be accomplished in different ways. For instance, we can simply upload a binary image to construct the source CSDA geometry 

We can now construct the final source CSDA by sampling the Gaussian Schell source model of equation ~(\ref{eq:GSM_source}). Once again, we have to compute all non-zero elements and can parallelize this task using PyOpenCL. Algorithm~\ref{alg:GSM} illustrates an example of how we can construct the cross-spectral density of a Gaussian Schell model-source.

\bibliography{PhDSpace_Ref}

\end{document}